\documentclass[twocolumn,english,amsart,showpacs,preprintnumbers,amsmath,amssymb,floatfix]{revtex4-1}
\usepackage{tikz,xcolor}
\usepackage[colorlinks = true,  
linkcolor = blue,
urlcolor  = blue,
citecolor = blue,
anchorcolor = blue]{hyperref}                    

\definecolor{lime}{HTML}{A6CE39}
\DeclareRobustCommand{\orcidicon}{%
	\begin{tikzpicture}
	\draw[lime, fill=lime] (0,0) 
	circle [radius=0.16] 
	node[white] {{\fontfamily{qag}\selectfont \tiny ID}};  
	\draw[white, fill=white] (-0.0625,0.095) 
	circle [radius=0.007];
	\end{tikzpicture}
	\hspace{-2mm}
}

\foreach \x in {A, ..., Z}{%
\expandafter\xdef\csname orcid\x\endcsname{\noexpand\href{https://orcid.org/\csname orcidauthor\x\endcsname}{\noexpand\orcidicon}}
}

\usepackage[T1]{fontenc}
\usepackage[latin9]{inputenc}
\usepackage{color}
\usepackage{array}
\usepackage{amstext}
\usepackage{graphicx}
\usepackage{esint}
\usepackage{rotating}
\usepackage{appendix}
\usepackage{float}

\usepackage[font=small,labelfont=bf]{caption}
\usepackage{xcolor}
\usepackage{ulem}

\makeatletter

\@ifundefined{textcolor}{}
{%
 \definecolor{BLACK}{gray}{0}
 \definecolor{WHITE}{gray}{1}
 \definecolor{RED}{rgb}{1,0,0}
 \definecolor{GREEN}{rgb}{0,1,0}
 \definecolor{BLUE}{rgb}{0,0,1}
 \definecolor{CYAN}{cmyk}{1,0,0,0}
 \definecolor{MAGENTA}{cmyk}{0,1,0,0}
 \definecolor{YELLOW}{cmyk}{0,0,1,0}
 }

\@ifundefined{definecolor}
 {\usepackage{color}}{}
\@ifundefined{definecolor}
 {\usepackage{color}}{}
\makeatother
\usepackage{babel}

\begin{document}


\title{ Nuclear parton distribution functions with uncertainties 
\\ 
in a general mass variable flavor number scheme }

\author{Hamzeh Khanpour$^{1,2,3}$\orcidA{}}
\email{Hamzeh.Khanpour@cern.ch}

\author{Maryam Soleymaninia$^{2}$\orcidB{}}
\email{Maryam\_Soleymaninia@ipm.ir}

\author{S. Atashbar Tehrani$^{2}$\orcidC{}}
\email{Atashbar@ipm.ir}

\author{Hubert Spiesberger$^{4}$\orcidD{}}
\email{spiesber@uni-mainz.de}

\author{Vadim Guzey$^{5}$\orcidE{}}
\email{guzey\_va@nrcki.pnpi.ru}

\affiliation {
$^{(1)}$Department of Physics, University of Science and Technology of Mazandaran, P.O.Box 48518-78195, Behshahr, Iran    \\
$^{(2)}$School of Particles and Accelerators, Institute for Research in Fundamental Sciences (IPM), P.O.Box 19395-5531, Tehran, Iran   \\
$^{(3)}$Department of Theoretical Physics, Maynooth University, Maynooth, Co. Kildare, Ireland  \\
$^{4}$PRISMA{\color{red}$^{+}$} Cluster of Excellence, Institut f\"ur Physik, Johannes-Gutenberg-Universit\"at, Staudinger Weg 7, D-55099 Mainz, Germany  \\
$^{5}$National Research Center ``Kurchatov Institute'', Petersburg Nuclear Physics Institute (PNPI), Gatchina 188300, Russia
}

\preprint{MITP/20-054}

\date{\today}

%
%

%
\begin{abstract}\label{abstract}

In this article we obtain a new set of nuclear parton distribution 
functions (nuclear PDFs) at next-to-leading order and 
next-to-next-to-leading order accuracy in perturbative QCD. 
The common nuclear deep-inelastic scattering (DIS) data analyzed 
in our study are complemented by the available charged-current 
neutrino DIS data with nuclear targets and data from Drell-Yan 
cross-section measurements for several nuclear targets. In addition, 
the most recent DIS data from the Jefferson Lab CLAS and Hall C 
experiments are also added to our data sample. For these specific 
datasets, we consider the impact of target mass corrections and 
higher twist effects which are expected to be important in the 
region of large $x$ and intermediate-to-low $Q^2$. 
Our analysis 
is based on a publicly available open-source tool 
{\tt APFEL}, 
which has been modified to be applicable for our analysis of 
nuclear PDFs. Heavy quark contributions to nuclear DIS are 
considered within the framework of the {\tt FONLL} general-mass 
variable-flavor-number scheme. The most recent {\tt CT18} PDFs 
are used as baseline proton PDFs. The uncertainties of nuclear 
PDFs are determined using the standard Hessian approach.
The results of our global QCD analysis are compared with existing 
nuclear PDF sets and with the fitted cross-sections, for which our 
set of nuclear PDFs provides a very good description.

\end{abstract}   
%


\pacs{12.38.-t, 24.85.+p, 13.15.+g, 13.60.Hb}

\maketitle
\tableofcontents{}

%
\section{Introduction}\label{sec:intro}

Nuclear parton distribution functions 
(nuclear PDFs)~\cite{AbdulKhalek:2019mzd,AbdulKhalek:2020yuc, 
Khanpour:2016pph,deFlorian:2011fp,Eskola:2009uj,Walt:2019slu, 
Kovarik:2015cma,Eskola:2016oht,Paukkunen:2020rnb,Kovarik:2015cma, 
AtashbarTehrani:2012xh,Kulagin:2014vsa,Eskola:2008ca,Kovarik:2010uv, 
Hirai:2001np,Hirai:2007sx,deFlorian:2012qw,Ru:2016wfx,Wang:2016mzo, 
Kusina:2016fxy} 
quantify the structure of quarks and gluons 
in nucleons bound in a nucleus and 
are essential ingredients for the calculation of hard 
scattering cross sections in charged-lepton deeply inelastic 
scattering (DIS) off nuclear targets and high-energy heavy-ion 
collisions. Based on the collinear factorization theorem, the 
non-perturbative nuclear PDFs are 
process independent and, as in the case of free proton PDFs, 
their scale dependence is governed by 
the standard DGLAP evolution equations~\cite{Dokshitzer:1977sg, 
Gribov:1972ri,Lipatov:1974qm,Altarelli:1977zs}. 
This framework has been shown to be consistent with 
experimental data on nuclear DIS and
heavy-ion collisions at the CERN-LHC.

A precise determination of nuclear PDFs is crucial for studies 
of the strong interaction in high-energy scattering processes in 
heavy-ion collisions, such as proton-lead (p-Pb) and lead-lead 
(Pb-Pb) collisions at the CERN-LHC. Furthermore, nuclear PDFs 
are important for high-energy neutrino interactions with 
heavy nuclear targets, which are sensitive to the 
separation of up- and down-type quarks, and hence, could provide 
important information for the decomposition of quark flavors 
in a QCD analysis~\cite{Walt:2019slu}. 

Several collaborations have recently presented new determinations 
of nuclear PDFs using the available experimental data, improved 
theoretical assumptions and advanced methodological settings. For 
the most recent determination of nuclear PDFs, we refer the reader 
to the analyses by the
{\tt nNNPDF}~Collaboration~\cite{AbdulKhalek:2019mzd, 
AbdulKhalek:2020yuc}, 
{\tt KA15}~\cite{Khanpour:2016pph}, 
{\tt EPPS16}~\cite{Eskola:2016oht}, 
{\tt TUJU19}~\cite{Walt:2019slu}, 
{\tt RKPZ}~\cite{Ru:2016wfx}, 
{\tt AT12}~\cite{AtashbarTehrani:2012xh}, 
{\tt KP14}~\cite{Kulagin:2014vsa}, 
{\tt DSSZ}~\cite{deFlorian:2011fp},    
{\tt HKN07}~\cite{Hirai:2007sx} and 
{\tt  nCTEQ15}~\cite{Kovarik:2015cma}.
Some of the mentioned nuclear PDF determinations are based on 
nuclear DIS data only. By using these data alone with a rather 
limited kinematic coverage, significant simplifying assumptions 
for the nuclear PDF parameterizations need to be taken into account. 
Hence, the constraints on the extracted quark and gluon nuclear 
PDFs are rather limited in these analyses. 

Nuclear PDFs at next-to-next-to-leading order (NNLO) accuracy 
in pQCD have been studied for the 
first time by {\tt KA15}~\cite{Khanpour:2016pph} 
in the zero-mass variable-flavor-number scheme (ZM-VFNS). 
The more recent work by 
{\tt nNNPDF1.0}~\cite{AbdulKhalek:2019mzd} is also performed at 
NNLO applying the NNPDF methodology and the resulting nuclear 
PDFs are determined by a Neural Network (NN) in the general-mass 
variable-flavor-number scheme (GM-VFNS). The most recent study 
by {\tt TUJU19}~\cite{Walt:2019slu} is performed at NNLO accuracy
as well, but 
based on the open-source {\tt xFitter} package~\cite{Alekhin:2014irh} 
within the nCTEQ framework. 

It is important to perform this study considering a different framework
and to define the bound  nucleon  
PDFs relative to a free nucleon baseline using the most recent proton PDF 
determination. The work presented in 
our paper focuses on the determination of new nuclear PDF sets, which we refer to as 
{\tt KSASG20}, at NLO and NNLO accuracy in pQCD. All available 
and up-to-date neutral current charged-lepton nuclear DIS,  
charged-current neutrino DIS experimental data, 
and the Drell-Yan (DY) cross-section ratios 
for several nuclear targets are used. 
The former two datasets are sensitive to the flavor 
composition.

Our analysis also incorporates the most recent DIS data for several 
nuclei at high $x$	from the Jefferson Lab CLAS and Hall C 
experiments. The JLab experiments provide a wealth of nuclear DIS 
data in the kinematic region of large Bjorken $x$ and intermediate 
to low photon virtuality $Q^2$. Hence, we particularly consider the 
impact of target mass corrections (TMCs) and higher twist effects 
(HT) which are expected to be important in the kinematic range of 
the JLab data.

The work presented in this paper is based on the publicly available 
open-source {\tt APFEL} package~\cite{Bertone:2013vaa}, which has 
been modified in order to accommodate the data from 
nuclear collisions, i.e., neutral-current charged lepton and 
charged-current neutrino DIS on nuclear targets. For the 
heavy-quark contributions, we use the 
{\tt FONLL-B} and {\tt FONLL-C} implementations of the GM-VFNS 
at NLO and NNLO, respectively. The standard 'Hessian' approach is 
used to estimate the uncertainties of nuclear PDFs for quarks 
and gluons due to experimental 
errors. The resulting uncertainties are examined 
in view of the sparse kinematic coverage of the available data.

For the free proton baseline, we use the most recent PDF analysis 
by {\tt CT18}~\cite{Hou:2019efy}, which is based mainly on the most 
recent measurements from the LHC and a variety of available world 
collider data. The {\tt CT18} PDFs are consistent with our 
assumptions and the kinematical cuts made for our nuclear PDF 
analysis. The nuclear PDFs presented in 
our study are available 
via the standard {\tt LHAPDF} library in order to provide an 
open-source tool for phenomenological applications. 

We should mention here that
a large amount of new and precise data from the LHC in proton-lead and lead-lead 
collisions became  recently
available~\cite{Chatrchyan:2014hqa, 
Khachatryan:2015hha,Khachatryan:2015pzs,Aad:2015gta, 
Sirunyan:2019dox}. These high precision data, especially 
the data on $W$ and $Z$ boson production in proton-lead collisions 
obtained by the ATLAS and CMS collaborations at center-of-mass energies of 
5.02~TeV and 8.16~TeV, could provide further constraints 
on nuclear PDFs, especially for the case of the nuclear gluon PDF. 
Their impact on nuclear PDFs has been extensively studied 
in Refs.~\cite{Eskola:2016oht,AbdulKhalek:2020yuc}.
In addition, an analysis of the impact of available
experimental data for proton-lead collisions
from Run I at the LHC on nuclear modifications of
PDFs is reported in Ref.~\cite{Armesto:2015lrg}
where the Bayesian reweighting 
technique~\cite{Eskola:2019dui,Paukkunen:2013grz,Paukkunen:2014zia} 
was used. 
In Ref.~\cite{Eskola:2019bgf} the impact of the single
inclusive $D^{0}$ meson production data
from LHCb~\cite{Aaij:2017gcy} in proton-lead collisions
on nuclear PDFs is quantified by the Hessian reweighting method.

It will be very interesting to repeat the analysis described 
here and determine a new set of nuclear PDFs by adding the data 
from proton-lead collisions at the LHC~\cite{Chatrchyan:2014hqa, 
Khachatryan:2015hha,Khachatryan:2015pzs,Aad:2015gta, 
Sirunyan:2019dox}.
In terms of future work, we plan to revisit this study and consider 
also the hadron collider data from the LHC.

The remainder of the paper is organized as follows:
In Sec.~\ref{sec:Theoretical-formalism}, we briefly review the 
general theoretical formalism for a global QCD analysis of nuclear 
PDFs and our assumptions for the input parameterization. This 
section also describes how we include target mass
corrections, higher twist effects, and  heavy flavor contributions 
in the nuclear PDF analysis. The charged lepton-nucleus DIS,  
very recent nuclear DIS data from 
JLab experiments, 
the neutrino(antineutrino)-nucleus DIS data, 
and the Drell-Yan cross-section measurements 
analyzed in our study 
are listed and discussed in Sec.~\ref{sec:Nucleardata}.
Then, in Sec.~\ref{sec:minimizations}, the procedure of $\chi^2$ 
minimization and the estimation of nuclear PDF uncertainties are 
presented. In Sec.~\ref{sec:FitResults}, we show and discuss in 
detail the global fit results and compare with other nuclear PDFs 
available in the literature. This section also includes our 
discussions of the fit quality and the data-theory comparison. 
Finally, the discussion and a summary of the main results 
are given in Sec.~\ref{sec:Discussion}.

%
\section{ Theoretical formalism and input distributions }\label{sec:Theoretical-formalism} 
%

This section presents the theoretical framework used in our 
analysis. First, we present the parameterization of the 
{\tt KSASG20} parton distributions of the nucleus. Then we 
discuss our method to include the heavy flavor contributions 
in the nuclear DIS processes.

\subsection{ The parton distributions of the nucleus }\label{sec:PDFs-nucleus}

In this section, we present our strategy to parameterize the 
{\tt KSASG20} nuclear PDFs at the input scale. 
Similarly to our 
previous analysis~\cite{Khanpour:2016pph} and as in other QCD 
analyses available in the literature~\cite{Hirai:2007sx, 
Eskola:2009uj}, we will work within the conventional approach 
which defines the nuclear PDFs, $x f_{i}^{N/A}(x,Q_0^2; A, Z)$, 
for a bound nucleon in a nucleus with the atomic mass number 
$A$ with respect to those for a free nucleon, 
$xf_{i}^{N}(x, Q_0^2)$, through a multiplicative nuclear 
modification factor, ${\cal W}_i(x, A, Z)$: 
\begin{eqnarray}\label{eq:nPDFs-Q0-1}
x f_{i}^{N/A}(x,Q_0^2; A, Z) = {\cal W}_i(x, A, Z) 
\times xf_{i}^{N}(x, Q_0^2) \,, 
\end{eqnarray}
where $i$ is an index to distinguish the distribution functions 
for the valence quarks $u_v$ and $d_v$, the sea-quarks $\bar{d}$, 
$\bar{u}$, the strange quark $s$ and $\bar{s}$, and the gluon $g$. 

Using the PDFs for a bound nucleon inside a nucleus presented in 
Eq.~\eqref{eq:nPDFs-Q0-1}, one can obtain the PDFs for a general 
nucleus $(A, Z)$ by averaging over the number of protons and 
neutrons inside the nuclei. It is given by,
\begin{eqnarray} 
\label{eq:nuclear}
f_{i}^{(A, Z)}(x, Q^{2}; A, Z) 
&&=  
\frac{1}{A} \left[ Z f_i^{p/A}(x, Q^{2}; A, Z) \right. 
\\  
&& \left. + (A-Z) f_i^{n/A}(x, Q^{2}; A, Z) \right] \,.
\nonumber
\end{eqnarray}
The bound neutron PDFs, $f_i^{n/A}(x, Q^{2}; A, Z)$, are obtained 
from the bound proton PDFs, $f_i^{p/A}(x, Q^{2}; A, Z)$, by assuming 
isospin symmetry in Eq.~\eqref{eq:nPDFs-Q0-1}. 

For the nuclear modification functions, ${\cal W}_i(x, A, Z)$, 
we follow the QCD analyses described in  
Refs.~\cite{Khanpour:2016pph,Paukkunen:2014nqa, 
AtashbarTehrani:2012xh,Hirai:2007sx,Eskola:2008ca, 
Rith:2014tma,Eskola:2012rg} and assume the following cubic-type 
modification function,
\begin{eqnarray}
\label{eq:weight-function}
{\cal W}_i(x, A, Z) &=&
1+\left(1-\frac{1}
{A^{\alpha}}\right)
\\
&& \times \frac{a_{i} (A)
+ b_{i}(A) \, x 
+ c_i(A) \, x^{2}
+ d_{i}(A) \, x^{3} }
{(1-x)^{\beta_{i}}}
\,. \nonumber
\end{eqnarray}
The advantage of the cubic form with the additional term $d_i$ in 
contrast to a quadratic-type function, i.e.\ with $d_i = 0$, is
that the nuclear modification becomes flexible enough 
to accommodate both shadowing and anti-shadowing in the valence 
quark distributions. For a detailed investigation of these 
functions, we refer the reader to 
Refs.~\cite{Khanpour:2016pph,Hirai:2001np,Hirai:2007sx}. 
 The explicit $A$-dependent pre-factor in 
${\cal W}_i$ in Eq.~\eqref{eq:weight-function} is constructed in 
such a way that for the proton ($A = 1$), one recovers the underlying 
free proton PDFs. The parameter $\alpha$ is considered to be fixed 
at $\alpha = \frac{1}{3}$. The two terms of the pre-factor 
$1 - A^{-\alpha}$ describe nuclear volume and surface 
contributions~\cite{Sick:1992pw,Hirai:2001np}. For the valence 
quark distributions, $\beta_v$ is fixed at 0.81, and for the sea 
quark and gluon densities, we fix $\beta_{\bar q}$ = $\beta_g$ 
= 1~\cite{Hirai:2001np}.

In general, all the coefficients $a_i$, $b_i$, $c_i$ and $d_i$ 
could carry an $A$- and a $Z$-dependence~\cite{Khanpour:2016pph}. 
Since the experimental data do not provide sufficient information 
to perform a stable fit for such a general ansatz, we therefore 
consider the $d$ parameter for the gluon density to be fixed at 
zero, i.e.,  $d_g = 0$. In order to give further and enough 
flexibility to our modification factor, we assume the following 
$A$-dependence of the parameters $a_i$ for sea-quarks, $b_i$ and 
$c_i$ for all parton flavors, and $d_i$ for valence and sea quarks,
\begin{eqnarray}
\label{eq:Adependent-function}
a_i(A)&=&a_i+(1-\frac{1}{A^{\epsilon _1}})~~~~(i=\bar{q}),\nonumber\\
b_i(A)&=&b_i+(1-\frac{1}{A^{\epsilon _2}})~~~~(i=v,\bar{q},g),\nonumber\\
c_i(A)&=&c_i+(1-\frac{1}{A^{\epsilon _3}})~~~~(i=v,\bar{q},g),\nonumber\\
d_i(A)&=&d_i+(1-\frac{1}{A^{\epsilon _4}})~~~~(i=v,\bar{q}).
\end{eqnarray}
The nuclear and neutrino DIS and Drell-Yan experiments analyzed in this study 
are performed on a variety of nuclear targets, and hence, allow us to constrain such an
$A$ dependence of our fit parameters. 
The coefficients $a_i (A)$ in 
Eq.~\eqref{eq:Adependent-function} depend on the atomic number $A$, 
but not all of them are free parameters that can be fitted. 
Among them only $a_{\bar q}$ needs to be determined from the 
QCD fit. There are three 
constraints for $a_{u_v}$, $a_{d_v}$ and $a_g$ due to the sum 
rules for the nuclear charge $Z$, the baryon number $A$, and 
from momentum conservation~\cite{Khanpour:2016pph,Hirai:2007sx}. 
The nuclear charge is given by
%
\begin{eqnarray} \label{eq:nuclear-charge}
Z = \int_{0}^{1} dx \, \frac{A}{3} 
\left[ 2 f_{u_v}^{(A,Z)} (x, Q_0^2) 
- f_{d_v}^{(A,Z)} (x, Q_0^2) \right] \, ,
\end{eqnarray}
the baryon number is expressed as 
\begin{eqnarray} \label{eq:Baryon_number}
A = \int_{0}^{1} dx \, \frac{A}{3} 
\left[ f_{u_v}^{(A,Z)} (x, Q_0^2) 
+ f_{d_v}^{(A,Z)} (x, Q_0^2) \right] \,,
\end{eqnarray}
and finally the momentum sum rule reads 
\begin{eqnarray} \label{eq:momentum-sum-rule}
\int_{0}^{1} dx  \sum_{i} \, x f_i^{(A,Z)} (x, Q_0^2) = 1 \,.
\end{eqnarray}

We emphasize that with these prescriptions the parameters 
$a_{i}$ for the valence up- and down-quark distributions 
are different, i.e.\ $a_{u_v} \neq  a_{d_v}$. 
Hence, the nuclear correction factors for the up- and down-valence 
distributions are not exactly the same, but expected to be 
similar in shape since the other parameters in the nuclear 
modification functions ${\cal W}_{u_v}$ and ${\cal W}_{d_v}$ 
are assumed to be the same. 
The remaining parameters in 
Eq.~\eqref{eq:weight-function} are obtained by a global 
$\chi^2$ analysis. 

In the {\tt KSASG20} nuclear PDF analysis, we use for the free 
proton PDFs the most recent PDF set of {\tt CT18}~\cite{Hou:2019efy} 
at the input scale $Q_0^2 = 1.69 \, \text{GeV}^2$, i.e., 
\begin{eqnarray}\label{eq:nPDFs-Q0-2}
xf_{i}^{p}(x, Q_0^2)  = xf_{i}^{p, \text{CT18}}(x, Q_0^2) \,. 
\end{eqnarray}
Considering the discussion above and following  
Eq.~\eqref{eq:nuclear}, the {\tt KSASG20} nuclear PDFs for all 
parton species can be explicitly written as~\cite{Khanpour:2016pph,Hirai:2007sx}: 
\begin{eqnarray} 
\label{eq:nuclear-PDFs}
 &&f_{u_v}^{(A, Z)}(x, Q_{0}^{2})= 
 \\ 
 && \quad \frac{1}{A}{\cal W}_{u_v}(x, A, Z)
    \left[Z\;f^p_{u_v}(x,Q_{0}^{2})+N\;f^p_{d_v}(x,Q_{0}^{2}) \right] \,,  
 \nonumber \\
 &&f_{d_v}^{(A, Z)}(x, Q_{0}^{2})= 
 \nonumber \\ 
 && \quad \frac{1}{A} {\cal W}_{d_v}(x, A, Z)
    \left[Z\;f^p_{d_v}(x, Q_{0}^{2})+N\;f^p_{u_v}(x, Q_{0}^{2})\right]\,, 
 \nonumber \\
 &&f_{\overline{u}}^{(A, Z)}(x, Q_{0}^{2})  =  
 \nonumber \\
 && \quad \frac{1}{A} 
    {\cal W}_{\overline{q}}(x, A, Z) 
    \left[Z\; f^p_{\overline{u}}(x, Q_{0}^2) 
    + N \, f^p_{\overline{d}}(x, Q_{0}^{2})\right] \, ,    
 \nonumber \\
 &&f_{\overline{d}}^{(A, Z)}(x,Q_{0}^{2}) = 
 \nonumber \\ 
 && \quad \frac{1}{A}  
    {\cal W}_{\overline{q}}(x, A, Z)
    \left[Z\; f^p_{\overline{d}}(x, Q_{0}^{2}) 
    + N \, f^p_{\overline{u}}(x, Q_{0}^{2}) \right] \, , 
 \nonumber \\
 &&f_{s}^{(A, Z)}(x, Q_{0}^{2}) = 
 f_{\overline{s}}^{(A, Z)}(x, Q_{0}^{2}) = 
 {\cal W}_{\overline{q}}(x, A, Z) \, f^p_{s}(x, Q_{0}^{2}) 
 \nonumber \,, \\
 &&f_g^{(A, Z)}(x, Q_{0}^{2})  =  
 {\cal W}_{g}(x, A, Z) \, f^p_g(x, Q_{0}^{2}) \,. 
 \nonumber
\end{eqnarray}
As one can see from our parameterizations,
we have assumed flavor dependent sea-quark densities, 
i.e., $f^{(A, Z)}_{\overline{d}} \neq f^{(A, Z)}_{\overline{u}}$, 
and the small differences between them  come from the
underlying free proton PDFs and the different number of protons and neutrons
in different nuclei.
The nuclear DIS data which we include in our analysis are not 
sensitive enough to constrain the sea-quark flavor decomposition, 
but the neutrino DIS data and the Drell-Yan cross-section measurements are 
sufficiently sensitive to the separation 
of up- and down-type quarks~\cite{Walt:2019slu,deFlorian:2011fp,Kovarik:2015cma,Eskola:2016oht}. 
Since we use {\tt CT18} as baseline 
proton PDFs, the strange quark distributions in nuclei are 
assumed to be flavor symmetric, $f^{(A, Z)}_{\overline{s}} 
=  f^{(A, Z)}_{s}$. We show that the parametrization presented 
above is sufficiently flexible to allow for
a good fit quality to the available datasets.

In our analysis, we define the momentum fraction $x$ with respect 
to the bound nucleon, $x=Q^2/[2 (q \cdot p_N)]$, where $q$ and 
$p_N$ are the photon and nucleon momenta, respectively.
With this convention, $x$ is allowed to vary in the interval 
$0<x<A$. However, recent studies available in the literature have 
shown that the nuclear structure functions fall off rapidly for
$x>1$~\cite{Fomin:2010ei,Niculescu:2005rh} and, hence,
can be neglected~\cite{Kovarik:2015cma,deFlorian:2011fp}. 
Therefore, to facilitate comparisons to other analyses, we 
follow the same path as other nuclear QCD analyses available 
in the literature and neglect the $x>1$ region.

\subsection{ Target Mass Corrections } 
\label{sec:Target-Mass-Corrections}

In this section, we describe in detail how we include target mass 
corrections in our analysis. TMCs are formally sub-leading 
$1 / Q^{2}$ corrections to the leading twist structure functions, 
where $Q^{2}$ is the squared four-momentum transfer to the hadron. 
TMC effects are most pronounced in the region of large-$x$ and 
moderate-to-small values of $Q^2$, which coincides with the region 
where nuclear PDFs are not very well determined from fits to the 
data. Since we include the neutral-current charged lepton DIS data 
from Jefferson Lab Hall C~\cite{Seely:2009gt} and 
CLAS~\cite{Schmookler:2019nvf} experiments, a reliable extraction 
of nuclear PDFs from these high-$x$ and low-$Q^{2}$ 
JLab data is expected to require an accurate consideration of TMCs in our QCD 
analysis.

The leading contributions to the TMC 
have been 
computed in Refs.~\cite{Georgi:1976ve,Schienbein:2007gr} and have 
been used in several studies in  the 
literature~\cite{Goharipour:2020gsw,Accardi:2016qay,Steffens:2012jx,Paukkunen:2020rnb,Moffat:2019qll,Khanpour:2017cha,VBertone-thesis}.
The target-mass corrected structure 
function, $F_{2}^{\rm TMC} (x, Q)$, is given by
\begin{eqnarray} 
\label{eq:TMC}
F_{2}^{\rm TMC} (x, Q) = 
\frac{x^2}{\xi^{2} 
\tau^{3}} F_{2}^{0} (\xi, Q) +   
\frac{6 \eta x^{3}}{\tau^2} 
\int_{\xi}^{1} dy 
F_{2} (y, Q)/y^{2} \,, 
\nonumber \\
\end{eqnarray}
where $\eta = M_p^2 / Q^2$ and $M_p$ is the mass of the proton.
In the above equation, $\xi = 2x / (1+\tau)$ refers to the 
Nachtmann variable~\cite{Nachtmann:1973mr}
with $\tau = 1 + 4 \eta x^2$, which 
shows that the TMCs vanish in the limit $M^2_p /Q^{2} \to 0$.
The superscript in $F_2^{0} (\xi, Q)$ indicates 
the limit, when  the proton mass $M_p$ is set to zero, 
and the second term in Eq.~\eqref{eq:TMC} 
gives the convolution term.
It is found that the magnitude of these corrections could be sizeable 
for lower values of the photon virtuality $Q \sim M_p$, and hence, 
one needs to take TMCs into account in the 
high-$x$ and low-$Q$ region in a QCD fit. 
In our analysis of the JLab data we include the leading 
target mass corrections due to using the Nachtmann variable 
and the convolution term proportional to $\eta$, but omit 
terms of order $O(\eta^2)$ which are also known~\cite{Schienbein:2007gr}.

\subsection{ Higher Twist Corrections } 
\label{sec:Higher-Twist-Corrections}

The inclusion of higher twist corrections is  
particularly important at high $x$ and low $Q^2$. 
To include these effects, we use the common 
phenomenological $x$-dependent function from the  
study by CJ15~\cite{Accardi:2016qay} and other studies available in the 
literature~\cite{Blumlein:2012bf,Accardi:2009br,Brady:2011uy,Goharipour:2020gsw}.
It is given by
\begin{eqnarray} 
\label{eq:HT}
F_{2}^{A} (x, Q) \to 
F_{2}^{(\rm LT, A)} (x, Q) 
[1 + C_{\rm HT} 
(x, A)/Q^{2}]\,, 
\end{eqnarray}
where 
$F_2^{(\rm LT, A)}$ indicates 
the leading twist structure function.
By assuming a simple $A^{1/3}$ scaling of HT effects 
for light nuclei, the function $C_{\rm HT} (x, A)$ can be 
written as~\cite{Qiu:2001hj},
\begin{eqnarray} 
\label{eq:CHT}
C_{\rm HT} (x, A) = H_0 ~ x^{H_1} 
(1 + H_2 x) A^{1/3} \, .
\end{eqnarray}
In fact, we will need this only for carbon. 
For the $H_i$ parameters included in 
$C_{\rm HT} (x, A)$, we use 
\{$H_0$, $H_1$, $H_2$\}=\{$-3.28$~GeV$^2$, 1.92, $-2.07$\} from CJ15~\cite{Accardi:2016qay} for our NLO analysis.
In addition, for our NNLO analysis, we follow the study presented in 
Ref.~\cite{Goharipour:2020gsw} and use 
\{$H_0$, $H_1$, $H_2$\}=\{$-1.32$~GeV$^2$, 1.49, $-1.96$\}.

\subsection{ Heavy flavor contributions } 
\label{sec:heavy-flavor}  

We note that the correct treatment of heavy quark mass 
contributions is important for global PDF analyses.
To account for the mass dependence in the {\tt KSASG20} nuclear 
PDF analysis for the charm and bottom PDFs, we treat 
heavy quarks within the GM-VFNS. We refer the reader to  
Refs.~\cite{Accardi:2016ndt,Martin:2009iq} for a detailed 
overview. 
For the {\tt KSASG20} analysis we use the heavy-quark 
schemes implemented in the public {\tt APFEL} 
package~\cite{Bertone:2013vaa}. At NLO we choose the scheme 
{\tt FONLL-B} which implements the NLO massive scheme 
calculation with NLO PDFs, while at NNLO we believe that 
the scheme {\tt FONLL-C} is the better choice since it 
combines the NLO massive scheme calculation with NNLO 
PDFs~\cite{Cacciari:1998it,Forte:2010ta}. We refer the 
reader to Ref.~\cite{VBertone-thesis} for more details of 
these schemes. 

The {\tt CT18} proton PDFs~\cite{Hou:2019efy}, which we 
consider as a baseline in our study, is based on the 
Aivazis-Collins-Olness-Tung (SACOT-$\chi$)
heavy-quark scheme~\cite{Kramer:2000hn,Tung:2001mv}, 
which is widely used in the CTEQ family of PDF fits. 
However, in our theoretical calculations and determination 
of nuclear PDFs, we use the {\tt FONLL} GM-VFN scheme. 
In order to examine the potential mismatch between these 
two schemes at NLO, we have calculated the inclusive DIS 
cross sections for a proton target using the SACOT-$\chi$ 
and {\tt FONLL-B} mass schemes. The results showed that 
the differences between these two schemes on the calculated 
cross sections are rather small, especially for $x>0.01$, 
i.e.\ the range which is covered by a large amount of nuclear 
data. In addition, the bulk of the data sets that we used in 
our study corresponds to cross-section ratios rather than to 
absolute cross-sections resulting in even smaller sensitivity 
to the choice of the scheme. Moreover, the heavy-quark 
production does not play a significant role in the inclusive 
cross sections at $x > 0.2$~\cite{Paukkunen:2020rnb}.

In summary, the choice of the heavy flavor and mass scheme
is not particularly critical for our analysis and we expect 
that it will have a negligible effect on the calculated 
cross sections in the kinematic regions covered by the nuclear 
DIS data. We refer the reader to the findings highlighted in
Ref.~\cite{Paukkunen:2020rnb} for detailed discussions.

In order to remain consistent with the 
{\tt CT18}~\cite{Hou:2019efy} baseline proton PDF 
analysis, for both the NLO and the NNLO fits, the heavy-quark 
masses are fixed at $m_c = 1.30$ GeV and $m_b$ = 4.75 GeV. 
The strong coupling constant is set equal to $\alpha_s (M_Z) 
=  0.118$~\cite{Tanabashi:2018oca} for both the NLO and 
NNLO fits.

\section{ Nuclear DIS datasets }
\label{sec:Nucleardata}

In this section, we discuss in detail the datasets which we have 
used in the {\tt KSASG20} nuclear PDF analysis. 
 
First, we start by presenting the neutral-current charged 
lepton-nucleus ($\ell^\pm A$) DIS data. These include the bulk 
of the datasets in our analysis which help to extract 
well-constrained valence and sea quark distributions;
however, they provide only a limited sensitivity to the 
gluon distribution and to the distinction of different  
quark flavors. Then we discuss the charged-current 
neutrino(antineutrino) DIS experimental data with nuclear 
targets. They depend on different combinations of quark flavors,  
compared with the neutral current case. The combination of 
neutral- and charged current data is, however, expected to 
be sufficiently sensitive to the flavor composition of 
non-isoscalar nuclei~\cite{Walt:2019slu}. 
The neutral-current charged lepton DIS data from Jefferson 
Lab Hall C and the most recent CLAS data measured by 
JLab will also
be discussed. Finally, we present the Drell-Yan dilepton pair 
production cross-section measurement by the Fermilab experiments 
E772 and E886, which are important to separate different quark 
flavors. This section also includes a detailed investigation of 
our data selection and kinematical cuts that need to be made when 
performing the {\tt KSASG20} global QCD analysis.

\subsection{ Neutral-current charged lepton nucleus DIS}
\label{sec:Nuclear-DIS-Data}

The neutral-current charged $\ell^\pm A$ DIS 
process is a powerful tool to study the nuclear structure and 
to extract 
nuclear PDFs. Hence, we consider the nuclear DIS 
data in the analysis as a baseline. The DIS of charged-leptons 
off nuclear targets, which initiated all studies of nuclear PDFs, 
provide the best constraints on nuclear modifications of 
the quark densities. 
These data are usually presented as a ratio 
of structure functions for two different nuclei and span the 
range from $0.005$ to $0.95$ in momentum fraction $x$ with a  
maximum photon virtuality of $Q^2_{\rm max} = 123$ GeV$^2$. 
The nuclear DIS data at lower momentum fractions, namely $x < 0.01$, 
are sensitive to the nuclear modifications of sea quarks, 
${\cal W}_{\bar q}$. The data at medium-to-large $x$ mainly 
probe the valence quark densities. A separation between 
quarks and antiquarks is not possible with these data alone. 
Other available data such as Drell-Yan dilepton production 
and (anti)neutrino collisions off nuclear targets should be 
used to provide a discrimination between valence and sea 
quarks. 

All available modern inclusive DIS measurements of 
neutral-current structure functions on nuclear targets are 
considered in our {\tt KSASG20} analysis. In particular, we use 
the nuclear DIS data from the NMC, EMC, and BCDMS experiments 
at CERN~\cite{Arneodo:1996ru,Benvenuti:1987az,Amaudruz:1995tq, 
Ashman:1988bf,Arneodo:1995cs,Bari:1985ga,Ashman:1992kv, 
Ashman:1988bf,Arneodo:1996rv}, measurements from 
SLAC~\cite{Arneodo:1989sy,Gomez:1993ri,Bodek:1983ec,Bodek:1983qn, 
Dasu:1988ru}, HERMES measurements at HERA~\cite{Ackerstaff:1999ac}, 
as well as the data from E665   
experiment at the Fermilab~\cite{Adams:1995is,Adams:1992nf}. 
The measurements of the nuclear structure functions in such 
experiments are typically 
presented as ratios of two different 
nuclei
\begin{eqnarray}\label{eq:Ratio}
R(x, Q^2; A_1, A_2)&=& \frac{F_2 (x, Q^2; A_1)}{F_2(x, Q^2; A_2)}\,.
\end{eqnarray}
In Tables~\ref{table:nuclear-DIS-data-D} and 
\ref{table:nuclear-DIS-data-C-Li}, the measured nuclear targets 
used in our {\tt KSASG20} QCD analysis are listed. For each 
dataset, we indicate the nuclei $A_1$ and $A_2$, which are used 
to construct the above structure function ratios. In addition, 
the experiments, the corresponding number of data points after 
cuts, and the published references are shown as well. In order 
to judge the quality of the fits, the values of $\chi^2$ 
extracted from our NLO and NNLO analyses are also presented.
As can be seen from Tables~\ref{table:nuclear-DIS-data-D} and 
\ref{table:nuclear-DIS-data-C-Li}, the number of available data 
points varies for different nuclei. A very large number of data 
points is available for the deuteron 
(Table~\ref{table:nuclear-DIS-data-D}) and for heavier nuclei, 
such as carbon (Table~\ref{table:nuclear-DIS-data-C-Li}). For 
other nuclei, such as e.g.\ lithium, only a few data points are 
available (Table~\ref{table:nuclear-DIS-data-C-Li}).

\begin{table*}
	\small
	\begin{center}
		\begin{tabular}{ c  c c c c c }
			\hline\hline {\tt Nucleus}     ~&~    {\tt Experiment}   ~&~   {\tt Number of data points}   ~&~ $\chi^2_{\tt NLO}$  ~&~ $\chi^2_{\tt NNLO}$ ~&~  {\tt Reference }   \\
			\hline\hline
			F$_2^A$/F$_2^D $   &     &    &   &    &   \\  \hline
			He/D &  {\tt SLAC-E139}   & 18 &     21.86 &  21.86 &   \cite{Arneodo:1989sy}     \\
			&  {\tt NMC-95}           & 16 &     9.91 &  9.84 &    \cite{Amaudruz:1995tq}    \\
			Li/D & {\tt NMC-95}      & 15 &      12.16 &  12.92 &    \cite{Amaudruz:1995tq}    \\
			Li/D\,(Q$^{2}$dep.)  & {\tt NMC-95} & 153 &   163.87  & 168.86 &  \cite{Arneodo:1995cs}   \\
			Be/D & {\tt SLAC-E139}   & 17 &       41.68 & 38.40  &   \cite{Gomez:1993ri}     \\
			C/D & {\tt EMC-88}       & 9 &       8.97 & 9.13 &    \cite{Ashman:1988bf}    \\
			& {\tt EMC-90}           & 2 &       0.13 & 0.05  &    \cite{Arneodo:1989sy}   \\
			& {\tt SLAC-E139}        & 7 &       14.56 & 14.05  &  \cite{Gomez:1993ri}     \\
			& {\tt NMC-95}           & 15 &      7.78 & 7.15 &   \cite{Amaudruz:1995tq}   \\
			& {\tt FNAL-E665 }       & 4 &        3.81 &  3.50  &   \cite{Adams:1995is}      \\  
			C/D\,(Q$^{2}$dep.) & {\tt NMC-95} & 164 &    144.90 &  146.02 &  \cite{Arneodo:1995cs}    \\
			N/D & {\tt BCDMS-85}     & 9 &      10.20  & 12.10 &    \cite{Bari:1985ga}      \\
			& {\tt HERMES-03}        & 92 &     55.72 &  65.12  &   \cite{Ackerstaff:1999ac}    \\  
			Al/D & {\tt SLAC-E49}    & 18 &     31.39  & 30.18  &  \cite{Bodek:1983ec}      \\
			& {\tt SLAC-E139 }       & 17 &     7.23 &  6.64  &   \cite{Gomez:1993ri}      \\
			Ca/D & {\tt EMC-90}      & 2 &      1.96 & 1.78  &   \cite{Arneodo:1989sy}    \\
			& {\tt NMC-95 }          & 15 &     30.91 & 39.48  &   \cite{Amaudruz:1995tq}  \\
			& {\tt SLAC-E139 }       & 7 &      4.28 & 4.02 &   \cite{Arneodo:1989sy}   \\
			& {\tt FNAL-E665 }       & 4 &       5.39 & 5.99 &    \cite{Adams:1995is}     \\
			Fe/D & {\tt SLAC-E87 }   & 14 &     7.18 &  7.61 &    \cite{Bodek:1983qn}     \\
			& {\tt SLAC-E139 }       & 23 &      27.58 & 25.92 &     \cite{Gomez:1993ri}     \\
			& {\tt SLAC-E140 }       & 6 &       10.69 & 10.93  &    \cite{Dasu:1988ru}      \\
			& {\tt BCDMS-87 }        & 10 &     16.60 & 15.74 &    \cite{Benvenuti:1987az}   \\
			Cu/D & {\tt EMC-93 }     & 19 &      12.15 &  12.59 &   \cite{Ashman:1992kv}      \\
			Kr/D & {\tt HERMES-03}   & 84 &     73.67 & 88.16  &   \cite{Ackerstaff:1999ac}    \\
			Ag/D & {\tt SLAC-E139}   & 7 &      11.12 & 14.47 &     \cite{Gomez:1993ri}     \\
			Sn/D & {\tt EMC-88}      & 8 &      16.85 & 18.72  &    \cite{Ashman:1988bf}   \\
			Xe/D & {\tt FNAL-E665-92} & 4 &     3.24 & 2.84  &    \cite{Adams:1992nf}     \\
			Au/D & {\tt SLAC-E139}   & 18 &     31.85 & 34.77  &    \cite{Gomez:1993ri}     \\
			Pb/D & {\tt FNAL-E665-95} & 4 &     9.01 & 8.64  &      \cite{Adams:1995is}     \\ \hline\hline
			\textbf{Total} &                   & \textbf{781} &            &   &   \\   \hline      
		\end{tabular}
		\caption[]{ 
			The charged lepton DIS experimental datasets for 
			$F_{2}^{A}$/$F_{2}^{D}$ used in the {\tt KSASG20} nuclear 
			PDF analysis. The specific nuclear targets, the experiment, 
			the number of data points, and the related references are listed. 
			The values of $\chi^2$ for the individual dataset obtained  
			in our NLO and NNLO fits are shown as well. 
		}
		\label{table:nuclear-DIS-data-D}
	\end{center}
\end{table*}

\begin{table*}   
	\small
	\begin{center}
		\begin{tabular}{c  c c c c c}
			\hline\hline {\tt Nucleus}     ~&~    {\tt Experiment}   ~&~   {\tt Number of data points}  ~&~ $\chi^2_{\tt NLO}$  ~&~ $\chi^2_{\tt NNLO}$ ~&~  {\tt Reference }   \\
			\hline\hline
			F$_2^A$/F$_2^C$   &     &    &   &   &         \\  \hline
			Be/C & {\tt NMC-96}     & 15 &  9.11  & 11.72 &        \cite{Arneodo:1996rv}    \\
			Al/C & {\tt NMC-96}     & 15 &   5.53 & 5.54 &       \cite{Arneodo:1996rv}    \\
			Ca/C & {\tt NMC-96}     & 20 &   14.85  & 13.82 &       \cite{Amaudruz:1995tq}   \\
			& {\tt NMC-96}          & 15 &  7.80 & 7.42  &         \cite{Arneodo:1996rv}   \\
			Fe/C & {\tt NMC-96}     & 15 &    9.04 &  8.50 &       \cite{Arneodo:1996rv}    \\
			Sn/C & {\tt  NMC-96}     & 144 & 135.86  & 150.24 &       \cite{Arneodo:1996rv}   \\
			& {\tt NMC-96}     & 15  &      21.51 &  27.21 &   \cite{Arneodo:1996ru}   \\
			Pb/C & {\tt NMC-96}     & 15 & 11.49 & 14.14 &       \cite{Arneodo:1996rv}    \\    		\hline
			\textbf{Total} &                   & \textbf{254} &  &  \\   \hline  \hline     
			F$_2^A$/F$_2^{Li}$   &     &    &   &   &         \\  \hline
			C/Li & {\tt NMC-95}     & 20 &   16.95  & 17.39 &      \cite{Amaudruz:1995tq}   \\
			Ca/Li & {\tt NMC-95}    & 20 &   23.49 & 25.34  &      \cite{Amaudruz:1995tq}     \\ \hline\hline
			\textbf{Total} &                   & \textbf{40} &  &   \\   \hline           
		\end{tabular} 
		\caption[]{
			Same as Table~\ref{table:nuclear-DIS-data-D}, but for the 
			charged lepton DIS datasets for $F_2^A$/$F_2^C$ 
			and $F_2^A$/$F_2^{Li}$. 
		}
		\label{table:nuclear-DIS-data-C-Li}   
	\end{center}
\end{table*}

In order to remain consistent with the {\tt CT18}~\cite{Hou:2019efy} 
baseline proton PDF analysis, we consider a kinematical cut on 
the momentum transfer $Q^2$: 
\begin{eqnarray}\label{eq:cut}
Q^2  \geq Q^2_{\rm min} = 1.69~{\rm GeV}^2 \,.
\end{eqnarray}
Our choice for the cut on $Q^2$ is the same as that of the 
{\tt EPS09}~\cite{Eskola:2009uj} and 
{\tt EPPS16}~\cite{Eskola:2016oht} nuclear PDF analyses. 
We do not impose any cut on the invariant square of the final 
state mass, $W^2$, again in agreement with what was done in 
Refs.~\cite{Eskola:2009uj,Eskola:2016oht}. After imposing the 
kinematical cut on $Q^2$ as presented in Eq.~\eqref{eq:cut}, we 
end up with a  total of $N_{\rm data} = 1075$ data points. 
As one can see from Table~\ref{table:nuclear-DIS-data-D}, a large 
amount of these points correspond to ratios of heavy nuclei with 
respect to deuterium.

In this work, we treat the deuteron as a nucleus. 
Hence, in addition to the nuclear DIS data discussed 
above, our analysis also includes the deuteron structure function 
$F_2^D$ measurements from NMC~\cite{Arneodo:1996qe}, 
BCDMS~\cite{Benvenuti:1989fm,Adams:1996gu}, 
HERMES~\cite{Airapetian:2011nu}, and finally the data for the 
deuteron-proton ratio $F_2^D/F_2^p$ from NMC~\cite{Arneodo:1996kd}. 
The deuteron data help to extract information on the 
flavor asymmetric antiquark distributions, $\bar d \neq \bar u$. 
Therefore, these data are essential for a successful QCD fit and 
for extracting information on the modification factors for the 
deuteron. For these specific datasets, a $Q^2 \geq 2 $ GeV$^2$ 
cut on the momentum transfer is considered.  
In Table~\ref{table:Deuteron-structure-function}, we list the 
measured deuteron structure function $F_2^D$ and deuteron-proton 
ratio $F_2^D/F_2^p$ used in the {\tt KSASG20} QCD analysis. After 
the cuts are applied, 529 data points remain, i.e.\ 373 for 
$F_2^D$ and 156 for $F_2^D/F_2^p$.

Compared with other analyses, we use a large amount of deuteron 
data in our fit.
As one can see from Table~\ref{table:Deuteron-structure-function}, 
the inclusion of 
higher-order QCD corrections significantly improves 
the description of {\tt NMC-96}~\cite{Arneodo:1996qe} 
and {\tt HERMES}~\cite{Airapetian:2011nu} data. A slight 
improvement is achieved for the case of 
{\tt BCDMS}~\cite{Benvenuti:1989fm} at NNLO, however, 
we obtained rather poor $\chi^2$ values both at NLO and NNLO. 
The $\chi^2$ values for the {\tt NMC-96}~\cite{Arneodo:1996kd} 
and the {\tt BCDMS} data from Ref.~\cite{Benvenuti:1989fm} 
deserve a separate comment. Despite of relatively good $\chi^2$ 
values for these specific datasets, the inclusion
of higher-order QCD correction does not improve the agreement 
of data with theory. 
The variation for the $\chi^2$ values that
can be seen in the table is possibly due to 
the tension among some of the datasets 
included in our QCD analysis, 
and the choice of input parameterization. 
It is important to note that two out of five deuteron-target data 
sets in Table~\ref{table:Deuteron-structure-function}, namely, 
the BCDMS $F_2^D$ data \cite{Benvenuti:1989fm} and the NMC-96 
$F_2^D/F_2^p$ data \cite{Arneodo:1996kd}, have already been 
used in the determination of the {\tt CT18} PDFs~\cite{Hou:2019efy} 
that we used as a baseline. This might result in certain double 
fitting of nuclear PDFs. We have checked that our {\tt KSASG20} PDFs 
and the {\tt CT18} PDFs lead to very similar predictions for the deuteron 
structure function $F_2^D(x,Q^2)$ in the kinematic range of the 
available data which shows the internal consistency of our 
approach. However, a more detailed study would be required to 
prove that the resulting uncertainties of our {\tt  KSASG20} nuclear 
PDFs are not underestimated by this double fitting issue.

\begin{table*}
	\small
	\begin{center}
		\begin{tabular}{c  c c c c c}
			\hline\hline {\tt Nucleus}     ~&~    {\tt Experiment}   ~&~   {\tt Number of data points}  ~&~ $\chi^2_{\tt NLO}$  ~&~ $\chi^2_{\tt NNLO}$ ~&~  {\tt Reference }   \\
			\hline \hline
			D & {\tt NMC-96}     & 126 &   201.90 &  88.50 &        \cite{Arneodo:1996qe}    \\
			D & {\tt BCDMS}     & 53 &    58.52 & 64.86 &       \cite{Adams:1996gu}   \\
			D & {\tt BCDMS}     & 155 &  245.89  &  216.84  &       \cite{Benvenuti:1989fm}    \\
			D & {\tt HERMES}          & 39 &  10.28  & 5.57 &    \cite{Airapetian:2011nu}   \\
			D/p & {\tt NMC-96}     & 156 & 148.98  & 152.71  &       \cite{Arneodo:1996kd}    \\ \hline 
			\textbf{Total} &                   & \textbf{529} &   &    \\   \hline              
		\end{tabular} 
		\caption[]{ 
			Same as Table~\ref{table:nuclear-DIS-data-D}, but for 
			data on
			the deuteron structure function $F_2^D$ and 
			the deuteron-proton ratio $F_2^D/F_2^p$. 
		}
		\label{table:Deuteron-structure-function}
	\end{center}
\end{table*}

\begin{figure*}[htb]
	\vspace{0.50cm}
	\resizebox{0.60\textwidth}{!}{\includegraphics{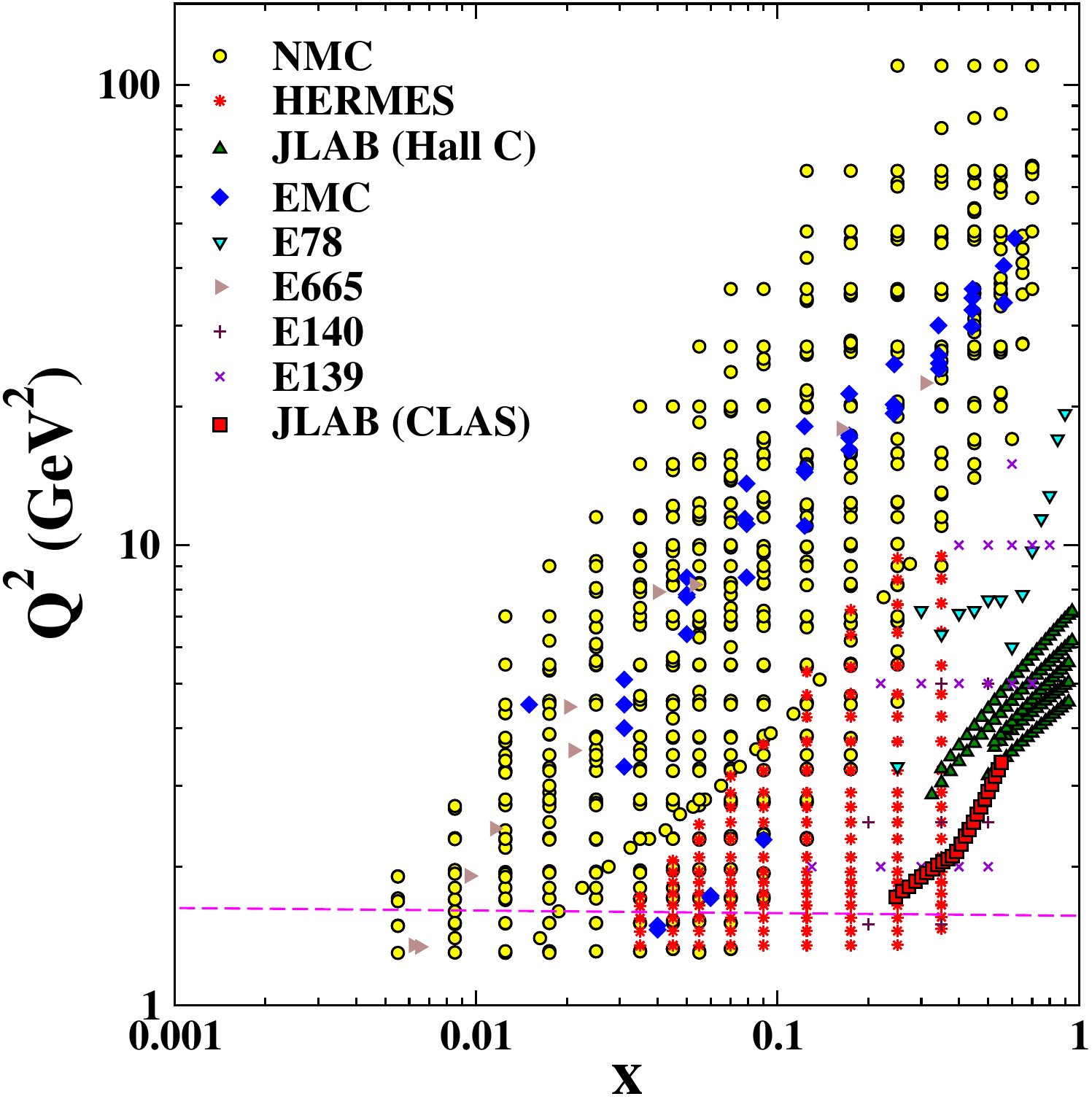}}
	\begin{center}
		\caption{{\small 
			The kinematic coverage of the world data for nuclear 
			DIS data used in the present global QCD analysis.
			The kinematic cut, which we apply in our fit, is 
			illustrated by the dashed line in the plot. 
			} 
			\label{fig:XQF2}}
	\end{center}
\end{figure*}

The kinematic coverage of the world data for nuclear DIS 
used in the present global QCD analysis is shown in 
Fig.~\ref{fig:XQF2} for some selected experiments. 
The applied kinematic cut on nuclear DIS data is illustrated 
by the dashed line in the plot.  The data points lying below 
the line are excluded in the present QCD analysis. 

\subsection{Nuclear DIS data from JLab experiments }
\label{sec:JLAB-Data}

JLab experiments have recently provided a wealth of nuclear 
DIS data in the region of large Bjorken $x$ and intermediate 
to low values of photon virtuality 
$Q^{2}$, which span a wide range
of $A$. These new JLab datasets that we consider in our analysis 
are for C, Al, Fe and Pb and they could provide high-precision 
constraints for the nuclear PDFs in the region of low 
$Q^{2}$ and 
high-$x$ (see Fig.~\ref{fig:XQF2}). The neutral-current charged 
lepton DIS experimental datasets for $F_{2}^{A}$/$F_{2}^{D}$ 
from Jefferson Lab Hall C~\cite{Seely:2009gt} and 
CLAS~\cite{Schmookler:2019nvf} measured by JLab during the 6~GeV 
electron beam operation~\cite{Gross:2011zz}	are listed in 
Table~\ref{table:nuclear-DIS-data-JLAB}.
In total, the number of data points from JLab is 199, 
among which 96 data points come from the CLAS collaboration.
	
\begin{table*}
\small
\begin{center}
\begin{tabular}{ c  c c c c c }
\hline\hline {\tt Nucleus}     ~&~    {\tt Experiment}   ~&~   {\tt Number of data points}   ~&~ $\chi^2_{\tt NLO}$  ~&~ $\chi^2_{\tt NNLO}$ ~&~  {\tt Reference }   \\
\hline\hline
F$_2^A$/F$_2^D $   &     &    &   &    &   \\  \hline
C/D     & {\tt JLAB Hall C}  & 103 &    158.88 & 154.51  &   \cite{Seely:2009gt}      \\
Pb/D 	& {\tt JLAB CLAS}     & 24 &    9.98 & 13.38  &   \cite{Schmookler:2019nvf}      \\
Fe/D 	& {\tt JLAB CLAS}     & 24 &    26.19 & 26.97  &   \cite{Schmookler:2019nvf}      \\
Al/D 	& {\tt JLAB CLAS}     & 24 &    16.39 & 15.23  &   \cite{Schmookler:2019nvf}      \\
C/D 	& {\tt JLAB CLAS}     & 24 &    14.19 & 14.66  &   \cite{Schmookler:2019nvf}      \\  \hline  \hline 
\textbf{Total} &                   & \textbf{199} &            &   &   \\   \hline             &
\end{tabular}
\caption[]{ 
	Neutral-current $\ell^\pm A$ DIS datasets for 
	$F_{2}^{A}$/$F_{2}^{D}$ from Jefferson Lab 
	Hall C~\cite{Seely:2009gt} and 
	CLAS~\cite{Schmookler:2019nvf} 
	at 6~GeV~\cite{Gross:2011zz}.
}
\label{table:nuclear-DIS-data-JLAB}
\end{center}
\end{table*}

In our NLO and NNLO QCD analyses of the JLab DIS data, we take 
into account the effect of TMCs as explained above in 
Sec.~\ref{sec:Target-Mass-Corrections}, and HT corrections 
as presented in Sec.~\ref{sec:Higher-Twist-Corrections}.
The values of $\chi^2$ for the individual data for different nuclei
extracted from our NLO and NNLO analyses are 
presented in Table.~\ref{table:nuclear-DIS-data-JLAB}.
As one can see, the $\chi^2$ values presented in this table 
reflect the fact that the new Hall C and 
CLAS data measured by JLab including the 
TMC and HT corrections are described 
well by our QCD fits. This agrees with the results of 
Ref.~\cite{Paukkunen:2020rnb}, where a PDF reweighting method 
is used to investigate the compatibility of the available 
nuclear PDFs studies, {\tt EPPS16}, {\tt TuJu19} and 
{\tt nCTEQ15}, with the recently published CLAS data.
Note, however, that global QCD fits, which explicitly 
include a particular dataset, e.g.\ the JLab data 
considered in our analysis, deliver more direct 
constraints on nuclear PDFs than the 
statistical reweighting of this data.

\subsection{ Charged-current (anti)neutrino nucleus DIS }
\label{sec:Neutrino-DIS-Data}

In addition to neutral-current DIS of charged leptons, we also 
include the data from neutrino-nucleus charged-current DIS 
experiments. Including the neutrino-nucleus DIS data in the 
analysis improves the nuclear PDF determination because 
it has additional sensitivity to the flavor composition of 
the PDFs due to the different couplings to down- and up-type 
quarks~\cite{Walt:2019slu}. 

In the {\tt KSASG20} nuclear PDF analysis, the cross section 
measurements of the CDHSW $\nu$ and $\bar{\nu}$ Fe 
experiment~\cite{Berge:1989hr} and the CHORUS  $\nu$ and 
$\bar{\nu}$ Pb experiment~\cite{Onengut:2005kv}  have been 
included. 

In the single-$W$ exchange approximation, 
the cross section for neutrino collisions with nuclear targets 
is given by 
\begin{eqnarray} 
\label{eq:neutrino-cross-section}
\frac{d^2\sigma^A}{dx dy} = && 
{\cal N} \left[\frac{y^2}{2} 2xF_1^{mA} 
+ \left(1 - y - \frac{M_Nxy}{2E_\nu}\right) 
F_2^{mA} \right.
\nonumber \\ 
&& \quad \quad
\left. \pm y \left(1 - \frac{1}{y}\right) xF_3^{mA} 
\right] \, , 
\end{eqnarray}
where $x$ and $y$ are the standard kinematical variables for a 
DIS process, $m=\nu$ and $\bar \nu$, and $A$ denotes the nucleus. 
The coupling factor ${\cal N}$  for (anti)neutrino-nucleus 
collisions can be written as 
\begin{eqnarray}\label{eq:N}
{\cal N} = 
\frac{G_F^2 M_N E_\nu}{\pi \left(1+\frac{Q^2}{M_W^2}\right)^2} 
\,. 
\end{eqnarray}
The cross section in Eq.~\eqref{eq:neutrino-cross-section} is 
described in terms of three structure functions, namely, $F_1^{mA}$, 
$F_2^{mA}$, and $F_3^{mA}$. 

The structure functions for neutrino scattering in 
Eq.~\eqref{eq:neutrino-cross-section} are given at leading order 
(LO) by~\cite{deFlorian:2011fp,Walt:2019slu,Tanabashi:2018oca}
\begin{eqnarray}\label{eq:neutrino-A}
F_2^{\nu A} &=&2x ( d^A + s^A + b^A + \bar u^A + \bar c^A  )\,, \nonumber \\ 
F_3^{\nu A} &=& 2( d^A + s^A + b^A - \bar u^A - \bar c^A )\,,
\end{eqnarray}
and for antineutrino scattering by 
\begin{eqnarray}
\label{eq:antineutrino-A}
F_2^{\bar \nu A} &=& 2x( u^A + c^A  + \bar d^A + \bar s^A + \bar b^A )\,, \nonumber \\ 
F_3^{\bar \nu A} &=& 2( u^A + c^A  - \bar d^A - \bar s^A - \bar b^A  )\,.
\end{eqnarray}
Full expressions including higher-order corrections 
can be found in Ref.~\cite{VBertone-thesis}.
As can be seen from Eqs.~\eqref{eq:neutrino-A} and 
\eqref{eq:antineutrino-A}, $F_2^{mA}$ is proportional 
to a particular non-singlet combination of quark distributions.
Hence, it is sensitive to both valence and sea quark densities.
In addition, $F_3^{mA}$ provides additional sensitivity to the flavor 
composition since it depends on a different linear combination 
of quark and antiquark PDFs. By combining the nuclear and neutrino 
DIS data, one can arrive at a considerably improved valence and sea 
quark separation in the entire region of $x$, where the data overlap.

In addition to the CDHSW and CHORUS neutrino DIS data, there are 
more neutrino scattering datasets available in the literature, 
namely the cross sections for an iron target measured by the 
NuTeV collaboration~\cite{Tzanov:2005kr} and the data from
the CCFR collaboration~\cite{MacFarlane:1983ax}.
For the CCFR
measurements, the quantities $Q^2$ and $x$ were 
not publicly available for the cross sections. In addition, only 
the averaged structure functions $F_{2}$ and $xF_{3}$ for
neutrino and anti-neutrino scattering on iron nuclei are 
available, which have less sensitivity to the flavor 
composition~\cite{Walt:2019slu}. Hence, we do not consider 
the CCFR data in our analysis.

Several studies in the literature have found some unresolved
tension between the NuTeV measurements and other lepton-nucleus 
data~\cite{Schienbein:2007fs,Schienbein:2009kk}.
A similar tension was also reported in 
Refs.~\cite{Kovarik:2010uv,MoosaviNejad:2016ebo} when
taking into account the neutrino
DIS data from the CHORUS and CCFRR measurements.
Detailed studies presented in 
Refs.~\cite{Paukkunen:2010hb,Paukkunen:2013grz} have 
shown that the tension with other data was specifically due 
to the inclusion of data from the NuTeV experiment.
Due to this tension, we have excluded the NuTeV neutrino 
DIS data from our QCD analysis.

\begin{figure*}[htb]
	\vspace{0.50cm}
	\resizebox{0.60\textwidth}{!}{\includegraphics{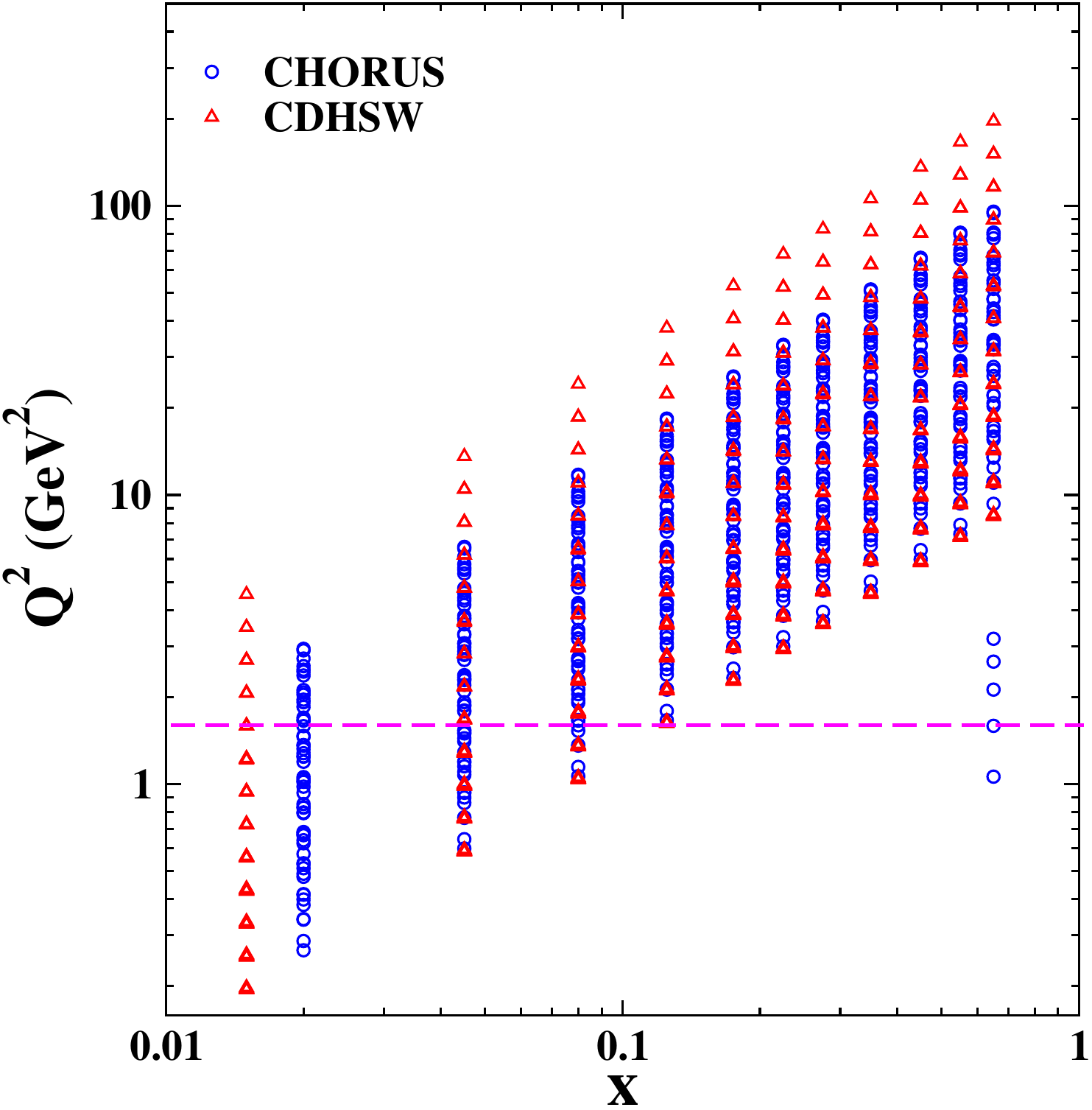}}
	\begin{center}
		\caption{{\small  
			The kinematic coverage of the neutrino DIS data used 
			in the present global QCD analysis. 
			Data points lying below the dashed line in the plot are 
			excluded from our fit. 
			} 
			\label{fig:XQSIGMA}}
	\end{center}
\end{figure*}

The $\nu A$ DIS data used in our analysis are presented in 
Table~\ref{table:neutrino-nucleus}. The number of data points, 
the respective reference and the specific nuclear target are 
listed as well. These datasets are subject to the same standard 
cut as presented in Eq.~\eqref{eq:cut}. In total, we include 
2458 data points for neutrino-nucleus collisions from CHORUS 
and CDHSW. The $\chi^2$ values obtained in our NLO and NNLO 
fits are shown as well. 
Figure~\ref{fig:XQSIGMA} shows the kinematic coverage of these 
data. Our kinematic cut is illustrated by the dashed line in the plot. 
The data points lying below the line are excluded from the present QCD 
analysis. 

\begin{table*}
\small
\begin{center}
\begin{tabular}{c  c c c c c}
			\hline\hline {\tt Nucleus}     ~&~    {\tt Experiment}   ~&~   {\tt Number of data points}  ~&~ $\chi^2_{\tt NLO}$  ~&~ $\chi^2_{\tt NNLO}$ ~&~  {\tt Reference }   \\ 	\hline\hline
			$\nu$ Pb & {\tt CHORUS}     & 532 &  459.71 & 569.92  &       \cite{Onengut:2005kv}   \\
			$\bar{\nu}$ Pb & {\tt CHORUS}    & 532 &  552.67  & 549.52 &      \cite{Onengut:2005kv} \\
			$\nu$ Fe & {\tt CDHSW}     & 698 & 790.06 & 733.92 &        \cite{Berge:1989hr}    \\
			$\bar{\nu}$ Fe & {\tt CDHSW}       & 696 &  695.79  & 679.00 &       \cite{Berge:1989hr}     \\ \hline\hline
			\textbf{Total} &                   & \textbf{2458} &   &   \\   \hline       
\end{tabular} 
\caption[]{
		The $\nu A$ DIS data used in the {\tt KSASG20} analysis. 
		The specific nuclear target, the experiment, the number of 
		data points, the values of $\chi^2$ extracted from our NLO 
		and NNLO fits, and the related references are listed. 
		}
\label{table:neutrino-nucleus}
\end{center}
\end{table*}

\subsection{ Drell-Yan cross-section data }\label{sec:Drell-Yan}

The analysis presented in our paper 
also incorporates Drell-Yan dilepton pair production.
The expression for the differential cross-section for 
the DY process is given in the 
literature~\cite{Hamberg:1990np,Kubar:1980zv,Song:2021mzt,Duan:2006hp,Mathews:2004xp,Hirai:2007sx,Sutton:1991ay}.
The proton-nucleus differential Drell-Yan cross-section 
can be written as a sum of two parts: the contributions 
of $q\bar{q}$ annihilation and the quark-gluon Compton process, 
\begin{eqnarray}
\label{eq:DY1}
\frac{d\sigma^{A}_{\text{DY}}}{dQ^2 dx_{F}} = 
\frac{d\sigma_{q \bar q}^{A}}{dQ^2 dx_{F}}
+\frac{d\sigma_{qg}^{A}}{dQ^{2} dx_F} .
\end{eqnarray}
Typically, the Drell-Yan cross-sections are measured in terms of the Feynman 
variable $x_{F}=x_{1}-x_{2}$,
where the variables $x_{1}$ and $x_{2}$ 
refer to the momentum fractions of the involved partons.
The subprocess cross-sections $d \hat{\sigma}_{q \bar{q}(0)}$
and $d \hat{\sigma}_{q \bar{q}(1)}$ for $q \bar{q}$ annihilation, 
and $d \sigma_{qg}$ for $qg$ and $\bar{q}g$ scattering, combined with 
PDFs are given by~\cite{Hirai:2007sx,Sutton:1991ay} 
\begin{eqnarray}
\label{eq:DY2}
&&\frac{d\sigma_{q\bar q}^{A}}{dQ^{2} dx_{F}} =  
\\ 
&& \quad
\frac{4 \pi \alpha^2}{9 Q^{2} s}
\sum_i e_i^{2} 
\int_{x_1}^1 dt_1 
\int_{x_2}^1 dt_2 
\bigg [ 
\frac{d \hat 
\sigma_{q\bar q(0)}}
{dQ^2 dx_F} 
+ 
\frac{d \hat \sigma_{q\bar q(1)}}
{dQ^{2} dx_{F}} \bigg ]
\nonumber \\
&& \quad 
\times 
\left[ q_i (t_1, Q^{2}) 
\bar q_i^A (t_2, Q^{2}) 
+ \bar q_i (t_1, Q^{2}) 
q_i^A (t_2, Q^{2}) \right],
\nonumber
\end{eqnarray}             
\begin{eqnarray}
\label{eq:DY3}
&&\frac{d\sigma_{qg}^{A}}{dQ^2 dx_F} = 
\\ 
&& \quad
\frac{4\pi \alpha^2}{9Q^2 s} 
\sum_i e_i^2 \int_{x_1}^{1}
dt_{1} \int_{x_2}^{1} dt_{2}
\nonumber \\
&& \quad
\times 
\bigg [ \frac{d\hat\sigma_{gq}}{dQ^2 dx_F}
g (t_1,Q^2) \left[ q_i^A 
(t_{2}, Q^2) + 
\bar q_i^{A} (t_2 ,Q^2) \right]
\nonumber \\
&& \quad \quad
+  \frac{d\hat\sigma_{qg}}{dQ^2 dx_F}
\left[ q_i (t_1, Q^2) + 
\bar q_i (t_1, Q^{2}) \right]
 g^A (t_2, Q^2)  \bigg ] .
\nonumber
\end{eqnarray}

As one can see from Eqs.~\eqref{eq:DY2} and \eqref{eq:DY3}, 
the Drell-Yan cross sections depend on the charged-weighted 
sum of all quark-antiquark flavors. This is again a different 
combination of quark flavors compared to the total inclusive 
DIS cross section. Therefore the combination of DIS and DY 
data may in principle help to separately constrain up and down 
quark distributions. 
We should stress here that Eqs.~\eqref{eq:DY2} and \eqref{eq:DY3} 
provide the DY cross section at NLO accuracy. For the calculation 
of the DY cross section at NNLO, we use a NNLO/NLO 
$K$-factor~\cite{Khanpour:2016pph}. In order to obtain this 
$K$-factor, we have computed the NNLO DY cross sections 
using the DYNNLO package~\cite{Catani:2007vq,Catani:2009sm} 
and then corrected the NLO cross sections applying this 
$K$-factor during the fit. 

The datasets for the Drell-Yan cross 
section ratios $\sigma_{\rm{DY}}^A/\sigma_{\rm{DY}}^{A^{\prime}}$ 
analyzed in the {\tt KSASG20} global QCD fit are listed in 
Table~\ref{table:Drell-Yan}. 
The specific nuclear targets, the number of data points, 
the extracted $\chi^2$ values for the NLO and 
NNLO QCD analysis, and the related 
references are listed in this table, as well.
In total, we include 92 DY data points.

\begin{table*}
\small
\begin{center}
\begin{tabular}{ c c c c c  c  }
\hline\hline {\tt Nucleus}     &     {\tt Experiment}   &   {\tt Number of data points}   ~&~ $\chi^2_{\tt NLO}$  ~&~ $\chi^2_{\tt NNLO}$ &    {\tt Reference}    \\
\hline\hline
$ \sigma_{\text{DY}}^A/\sigma_{\text{DY}}^{A^{\prime}} $  &  &   &     &     &    \\ \hline
Fe/Be & {\tt FNAL-E866/NuSea}   & 28 & 28.38 &  28.13   &  \cite{Vasilev:1999fa}    \\
W/Be & {\tt FNAL-E866/NuSea}    & 28 & 37.17  &  32.09   &  \cite{Vasilev:1999fa}    \\
C/D & {\tt FNAL-E772-90}   & 9 &  30.12  &  33.83   &   \cite{Alde:1990im}    \\
Ca/D & {\tt FNAL-E772-90}  & 9 & 4.35    & 6.13    &  \cite{Alde:1990im}    \\
Fe/D & {\tt FNAL-E772-90}  & 9 & 25.98   & 29.31  &   \cite{Alde:1990im}    \\
W/D & {\tt FNAL-E772-90}   & 9 & 14.04    & 14.44    &  \cite{Alde:1990im}    \\ \hline \hline
\textbf{Total}  &                   & \textbf{92} &  \\   \hline
\end{tabular}
\caption[]{  
	The Drell-Yan cross section ratios $\sigma_{{\text{DY}}}^A /
	\sigma_{{\text{DY}}}^{A^{\prime}}$ measured at FNAL used in 
	the {\tt KSASG20} nuclear PDFs global fit. The specific 
	nuclear target, the number of analyzed data points, and 
	the related reference are listed, as well. 
}
\label{table:Drell-Yan}
\end{center}
\end{table*}

\section{ $\chi^2$ minimization and uncertainty estimation }
\label{sec:minimizations}

The optimal values for the nuclear PDF parameters defined in 
Eq.~\eqref{eq:nuclear-PDFs} are extracted from the nuclear 
DIS and  and Drell-Yan data using the global function 
$\chi^{2}_{\text{global}}(\{\xi_{i}\})$ given by 
\begin{equation}
\chi_{\text{global}}^2(\{\xi_{i}\})
= \sum_{m=1}^{m^{\text{exp}}} \,
w_{m}  \,
\chi_m^2 (\{\xi_{i}\})\,,
\label{eq:chi2-1}
\end{equation}
where $m$ labels the experiment. $w_{m}$ allows us, in principle, 
to include datasets with different weight factors. However, we 
always use the default value $w_{m} = 1$ for all experimental 
datasets~\cite{Blumlein:2006be,Stump:2001gu,James:1975dr,Pumplin:2000vx,Salajegheh:2019ach,Soleymaninia:2018uiv}. 
Each experiment 
contributes with $\chi_m^2(\{\xi_{i}\})$ to the global $\chi^2$. 
These terms epend on the fit parameters $(\{\xi_{i}\})$, which 
are identified with the parameters of the bound proton PDFs at 
the initial scale. $\chi_m^2(\{\xi_{i}\})$ is calculated as 
\begin{eqnarray}
\chi_m^2(\{\xi_{i}\}) &=&
\left(\frac{1-{\cal N}_m}
{\Delta{\cal N}_m}\right)^2 +
\nonumber \\
\hspace{-0.5cm}\sum_{j=1}^
{N_m^{\text{data}}} &&
\left(\frac{({\cal N}_m  \,
{\cal O}_j^{\text{data}}-
{\cal T}_j^{\text{theory}}
(\{\xi_i\})}{{\cal N}_m \,
\Delta_j^{\text{data}}}\right)^2 \,. 
\label{eq:chi2-2}
\end{eqnarray}
Here $j$ runs over data points, $m$ indicates a given individual 
dataset, and $N^{\text{data}}_m$ corresponds to the 
total number of data points in this set. In the above equation, 
${\cal O}^{\text{data}}_{j}$ is the value of the measured data 
point for a given observable, and $\Delta^{\text{data}}_j$ 
is the experimental error calculated from the statistical and 
systematic errors added in quadrature. The theoretical predictions 
for each data point $j$ are represented by 
${\cal T}^{\text{theory}}_{j}(\{\xi_{i}\})$, which has to be 
calculated at the same experimental kinematic point $x$ and 
$Q^{2}$ using the DGLAP-evolved nuclear PDFs with given parameters 
$(\{\xi_{i}\})$. We use the CERN subroutine 
{\tt MINUIT}~\cite{James:1975dr} to determine the independent 
fit parameters of nuclear PDFs  $f_{i}^{(A, Z)}(x, Q^{2}; A, Z)$ 
by minimizing $\chi_{\text{global}}^2(\{\xi_{i}\})$. 

In Eq.~\eqref{eq:chi2-2}, ${\Delta{\cal N}_m}$ describes 
the overall normalization uncertainty for each charged
lepton DIS experiment. We include the normalization ${\cal N}_m$ 
of different experiments as a free parameter along with other 
independent fit parameters $(\{\xi_{i}\})$. 
First, we determine their values in a global pre-fit, 
then we fix them on their best-fitted values when 
we determine the uncertainties of the nuclear PDF parameters.  

The quality of the QCD fit can be estimated from the resulting 
$\chi^2 / N^{\text {data}}$, where $N^{\text {data}}$ indicates 
the number of data points.

After describing the method to obtain the central value of 
the {\tt KSASG20} nuclear PDFs by minimizing the 
$\chi_{\text{global}}^2(\{\xi_{i}\})$ function, we are in a 
position to present our method to estimate the uncertainties 
of our nuclear PDFs. There are three established methods, 
namely the Hessian method~\cite{Pumplin:2001ct,Pumplin:2000vx}, 
the Monte Carlo (MC) 
method~\cite{Hartland:2019bjb,AbdulKhalek:2019mzd} and Lagrange 
multiplier method~\cite{Stump:2001gu}, which can be used for the 
error analysis. The analysis of the uncertainties in {\tt KSASG20} 
is done using the standard 'Hessian' approach~\cite{Martin:2009iq, 
Pumplin:2000vx,Pumplin:2001ct,Eskola:2016oht}, which we will 
briefly describe in the following. For the uncertainty estimate, 
we follow the notation adopted in 
Refs.~\cite{Pumplin:2001ct,Pumplin:2000vx} and refer the reader 
to these publications for a detailed discussion of the Hessian 
formalism. 

The {\tt KSASG20} nuclear PDF uncertainties are estimated by 
using the Hessian matrix, $H$, defined by 
\begin{eqnarray}
\label{eq:uncertainties}
&&\left[\delta f^{(A,Z)}\right]^2 =
\\
&&T^2 \sum_{m,n} 
\left(\frac{\partial 
f^{(A,Z)}(x,\{\xi\})}
{\partial \xi_{m}}  
\right)_{\xi=
\hat{\xi}}
H_{mn}^{-1}
\left(\frac{\partial 
f^{(A,Z)}(x,\{\xi\})}
{\partial \xi_{n}} 
\right)_{\xi =
\hat{\xi}} \,, 
\nonumber
\end{eqnarray}
where $H_{mn}$ corresponds to the components of the Hessian matrix 
obtained by the CERN subroutine MINUIT, $\{\hat\xi\}$ indicates 
the set of optimum independent fit parameters, and $\xi_n$ are 
the fit parameters of the chosen functional form at the initial 
scale. The value of $T^2 = \Delta \chi_{\text{global}}^2$ in 
Eq.~\eqref{eq:uncertainties} is the tolerance for the required 
confidence interval. It is calculated so that the confidence 
level (CL) becomes the one-$\sigma$ error range, i.e.\ 68\% CL, 
for a given number of independent fit parameters. In an ideal 
case, with the standard `parameter-fitting' criterion for one 
free parameter, one would 
choose the tolerance criterion $T^2=\Delta \chi_{\text{global}}^2 
= 1$ for a 68\%, i.e.\ one-sigma CL, or $T^2 = 2.71$ for 
a 90\% CL~\cite{Martin:2009iq}. In the {\tt KSASG20} 
nuclear PDF analysis, the tolerance for $\chi_{\text{global}}^2$ 
is based on the method presented in Refs.~\cite{Eskola:2016oht, 
Martin:2009iq,Hirai:2007sx}. In our study with 18 free fit 
parameters it becomes $\Delta \chi_{\text{global}}^2 = 20$ at the 68\% CL. 

We note that other groups, e.g., 
{\tt nCTEQ15}, {\tt EPPS16} and {\tt TUJU19},
base their results on different values of the tolerance criterion:
$\Delta \chi^2 = 35$ for {\tt nCTEQ15}, 
$\Delta \chi^2 = 52$ for {\tt EPPS16}, and
$\Delta \chi^2 = 50$ for {\tt TUJU19}. 

%
%
\begin{table}
	\begin{center}
		\caption{\small 
			Best fit parameters and uncertainties of the 
			{\tt KSASG20} fits at NLO and NNLO at the initial scale 
			$Q_0^2 = 1.69$~GeV$^2$. Values marked with (*) are 
			fixed in our fit. 
		}
		\begin{tabular}{ c | c | c }
			\hline \hline
			Parameters	  & {\tt NLO}             & {\tt NNLO}           \\  \hline \hline
			$a_{v}$       & see Table.~\ref{tab:au-ad-ag-NLO}                 & see Table.~\ref{tab:au-ad-ag-NNLO}                \\
			$b_{v}$       & $0.350 \pm 0.05$  & $0.047\pm 0.040$  \\
			$\epsilon_{b_{v}}$       & $0.014\pm 0.005$  & $0.011 \pm 0.005$  \\
			$c_{v}$       & $-2.112 \pm 0.121$  & $-1.373 \pm 0.084$   \\
			$\epsilon_{c_{v}}$       & $-0.0005 \pm 0.008$  & $0.0001 \pm 0.008$   \\
			$d_{v}$       & $1.961 \pm 0.081$            & $1.520 \pm 0.057$           \\
			$\epsilon_{d_{v}}$       & $-0.007 \pm 0.007$  & $-0.006 \pm 0.008$  \\
			$\beta _{v}$  & $ 0.81^*$            & $ 0.81^*$           \\     \hline
			$a_{\overline{q}}$    & $-0.240\pm 0.011$  & $-0.210\pm 0.011$   \\
			$\epsilon_{a_{\overline{q}}}$    & $-0.005 \pm 0.001$  & $-0.011 \pm 0.001$   \\
			$b_{\overline{q}}$    & $4.731 \pm 0.254$  & $5.518\pm 0.271$   \\
			$\epsilon_{b_{\overline{q}}}$    & $0.228 \pm 0.037$  & $0.176 \pm 0.030$   \\
			$c_{\overline{q}}$    & $-23.594 \pm 1.673$              & $-25.086 \pm 1.715$              \\
			$\epsilon_{c_{\overline{q}}}$    & $0.154\pm 0.072$  & $0.121 \pm 0.064$   \\
			$d_{\overline{q}}$    & $29.061 \pm 2.667$           & $31.678 \pm 3.063$       \\
			$\epsilon_{d_{\overline{q}}}$       & $0.060 \pm 0.102$  & $0.039 \pm 0.104$  \\
			$\beta _{\overline{q}}$      & $1.0^*$               & $1.0^*$               \\   \hline
			$a_{g}$         & see Table.~\ref{tab:au-ad-ag-NLO}        & see Table.~\ref{tab:au-ad-ag-NNLO}   \\
			$b_{g}$         & $-2.400 \pm 0.456$               & $0.497 \pm 0.450$               \\
			$\epsilon_{b_{g}}$         & $0.174 \pm 0.084$               & $0.100 \pm 0.072$               \\
			$c_{g}$         & $6.433\pm 0.851$            & $1.786 \pm 0.663$        \\
			$\epsilon_{c_{g}}$         & $0.168 \pm 0.138$            & $-0.063 \pm 0.046$        \\
			$d_{g}$         & $0.0^*$  & $0.0^*$        \\ 	
			$\epsilon_{d_{g}}$& $0.0^*$  & $0.0^*$        \\ 			
			$\beta _{g}$    & $1.0^*$               & $1.0^*$                     \\ 					 \hline  \hline
			$\alpha_s(M_Z^2)$   & $0.118^*$ & $0.118^*$   \\
			$m_c$               & $1.30^*$   & $1.30^*$     \\
			$m_b$               & $4.75^*$   & $4.75^*$     \\ 	\hline \hline
			$\chi^2/{\rm d.o.f}$  & $4582.066/4335=1.056$   & $4548.630/4335=1.049$  \\  	\hline \hline
		\end{tabular}
		\label{tab-nuclear-pdf}
	\end{center}
\end{table}
%
%
%
%
\begin{table}
	\begin{center}
		\caption{ 
			Values of the parameters $a_{u_v}$, $a_{d_v}$, and $a_g$ 
			for several nuclei analyzed in this study at NLO 
			accuracy. These parameters are obtained from the sum 
			rules for the nuclear charge $Z$, the baryon number 
			$A$, and momentum conservation. See 
			Sec.~\ref{sec:PDFs-nucleus} for details. 
		}
		\begin{tabular}{ c || c  c  c }
			\hline    \hline
			Nucleus ~&~ $a_{u_v}$ ~&~ $a_{d_v}$ ~&~ $a_g$          \\    \hline   \hline
			$^2$D     ~&~  0.0111404  ~&~  0.0111404  ~&~ 0.0870223 \    \\
			$^4$He    ~&~ 0.00952477  ~&~  0.00952477  ~&~ 0.0560934 \    \\
			$^7$Li    ~&~ 0.00774508  ~&~  0.00867554  ~&~  0.0337899 \    \\
			$^9$Be    ~&~ 0.00728751  ~&~  0.00799997  ~&~ 0.0244899 \    \\
			$^{12}$C     ~&~  0.00700116  ~&~  0.00700116  ~&~  0.0143500 \    \\
			$^{14}$N     ~&~ 0.00665067  ~&~  0.00665067  ~&~ 0.00912677 \    \\
			$^{27}$Al    ~&~  0.00505471  ~&~  0.00527678  ~&~  $-$0.0115712 \    \\
			$^{40}$Ca    ~&~  0.00428701  ~&~  0.00428701  ~&~  $-$0.0228351 \    \\
			$^{56}$Fe    ~&~  0.00332813  ~&~  0.0037376  ~&~  $-$0.0318663 \    \\
			$^{63}$Cu    ~&~  0.00304467  ~&~  0.00349631  ~&~  $-$0.0349004 \    \\
			$^{84}$Kr    ~&~  0.00222106  ~&~  0.0030206  ~&~  $-$0.0420497 \    \\
			$^{108}$Ag    ~&~ 0.00171484  ~&~ 0.00242832  ~&~  $-$0.0479941 \    \\
			$^{119}$Sn    ~&~  0.00141338  ~&~  0.00228757  ~&~  $-$0.0502205 \    \\
			$^{131}$Xe    ~&~  0.00115708  ~&~  0.00211297  ~&~  $-$0.0523868 \    \\
			$^{197}$Au    ~&~  0.000212695  ~&~  0.00126285  ~&~  $-$0.0611757 \    \\
			$^{208}$Pb    ~&~  0.000055209  ~&~  0.00117403  ~&~  $-$0.0622995 \    \\    \hline   \hline
		\end{tabular}
		\label{tab:au-ad-ag-NLO}
	\end{center}
\end{table}
%
%
%
%
\begin{table}
	\begin{center}
		\caption{ 
			Same as Table.~\ref{tab:au-ad-ag-NLO}, but at NNLO 
			accuracy. 
		}
		\begin{tabular}{ c || c  c  c }
			\hline     \hline
			Nucleus ~&~ $a_{u_v}$ ~&~ $a_{d_v}$ ~&~ $a_g$ \\        \hline \hline
			$^2$D     ~&~  0.0332173  ~&~  0.0332173  ~&~  $-$0.245614 \    \\
			$^4$He    ~&~  0.0319167  ~&~ 0.0319167  ~&~  $-$0.258804 \    \\
			$^7$Li    ~&~  0.0303139  ~&~ 0.0313825  ~&~  $-$0.268223 \    \\
			$^9$Be    ~&~  0.0299825  ~&~  0.0308045  ~&~  $-$0.272104 \    \\
			$^{12}$C     ~&~  0.0298786  ~&~  0.0298786  ~&~  $-$0.276295 \    \\
			$^{14}$N     ~&~  0.0295949  ~&~ 0.0295949  ~&~  $-$0.278435 \    \\
			$^{27}$Al    ~&~  0.0282598  ~&~  0.0285215  ~&~  $-$0.286744 \    \\
			$^{40}$Ca    ~&~  0.0276775  ~&~  0.0276775  ~&~  $-$0.291109 \    \\
			$^{56}$Fe    ~&~  0.0268174  ~&~  0.0273072  ~&~  $-$0.294503 \    \\
			$^{63}$Cu    ~&~ 0.0265775  ~&~  0.0271192  ~&~  $-$0.295617 \    \\
			$^{84}$Kr    ~&~  0.0258313  ~&~  0.0267962  ~&~  $-$0.298186 \    \\
			$^{108}$Ag    ~&~  0.0254344  ~&~  0.0263003  ~&~  $-$0.300247 \    \\
			$^{119}$Sn    ~&~ 0.025152  ~&~  0.0262153  ~&~  $-$0.301002 \    \\
			$^{131}$Xe    ~&~  0.0249232  ~&~  0.0260883  ~&~  $-$0.301725 \    \\
			$^{197}$Au    ~&~ 0.0241236  ~&~  0.0254157  ~&~  $-$0.304532 \    \\
			$^{208}$Pb    ~&~  0.0239777  ~&~  0.025356  ~&~  $-$0.304877 \    \\    \hline \hline
		\end{tabular}
		\label{tab:au-ad-ag-NNLO}
	\end{center}
\end{table}
%
%

\section{ Results and discussions }\label{sec:FitResults}   

In the following section we present the main results and findings 
of our QCD analysis. We first discuss the main features of the 
{\tt KSASG20} nuclear PDF parameters. Then, we assess the stability 
of our NLO and NNLO results with respect to the perturbative order. 
We present a detailed comparison with the recent NLO and NNLO 
nuclear PDF analyses available in the literature. Finally, the 
section is concluded with a discussion of the quality of our fit 
results by comparing the resulting structure function ratios with 
the nuclear DIS experimental data; we compare our theoretical 
predictions with the neutrino DIS and Drell-Yan data as well.

\begin{figure*}[hb!]
	\vspace{0.50cm}
	\resizebox{0.990\textwidth}{!}{\includegraphics{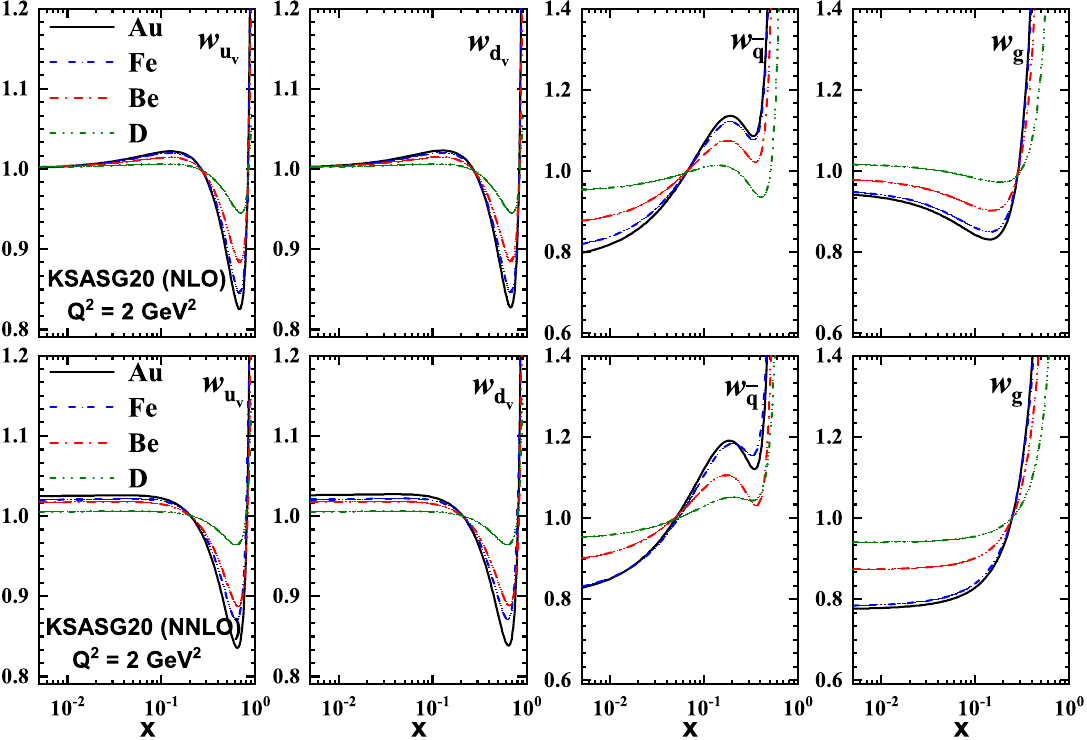}}
	\begin{center}
		\caption{{\small 
			The nuclear modification factors for deuterium (D), Beryllium (Be), 
			iron (Fe) and gold (Au) at NLO (top row) 
			and NNLO (bottom row) accuracy in pQCD and at the 
			scale $Q^2$ = 2 GeV$^2$.
			} 
			\label{fig:w-NLO-NNLO}}
	\end{center}
\end{figure*}
\begin{figure*}[htb]
	\vspace{0.50cm}
	\resizebox{0.990\textwidth}{!}{\includegraphics{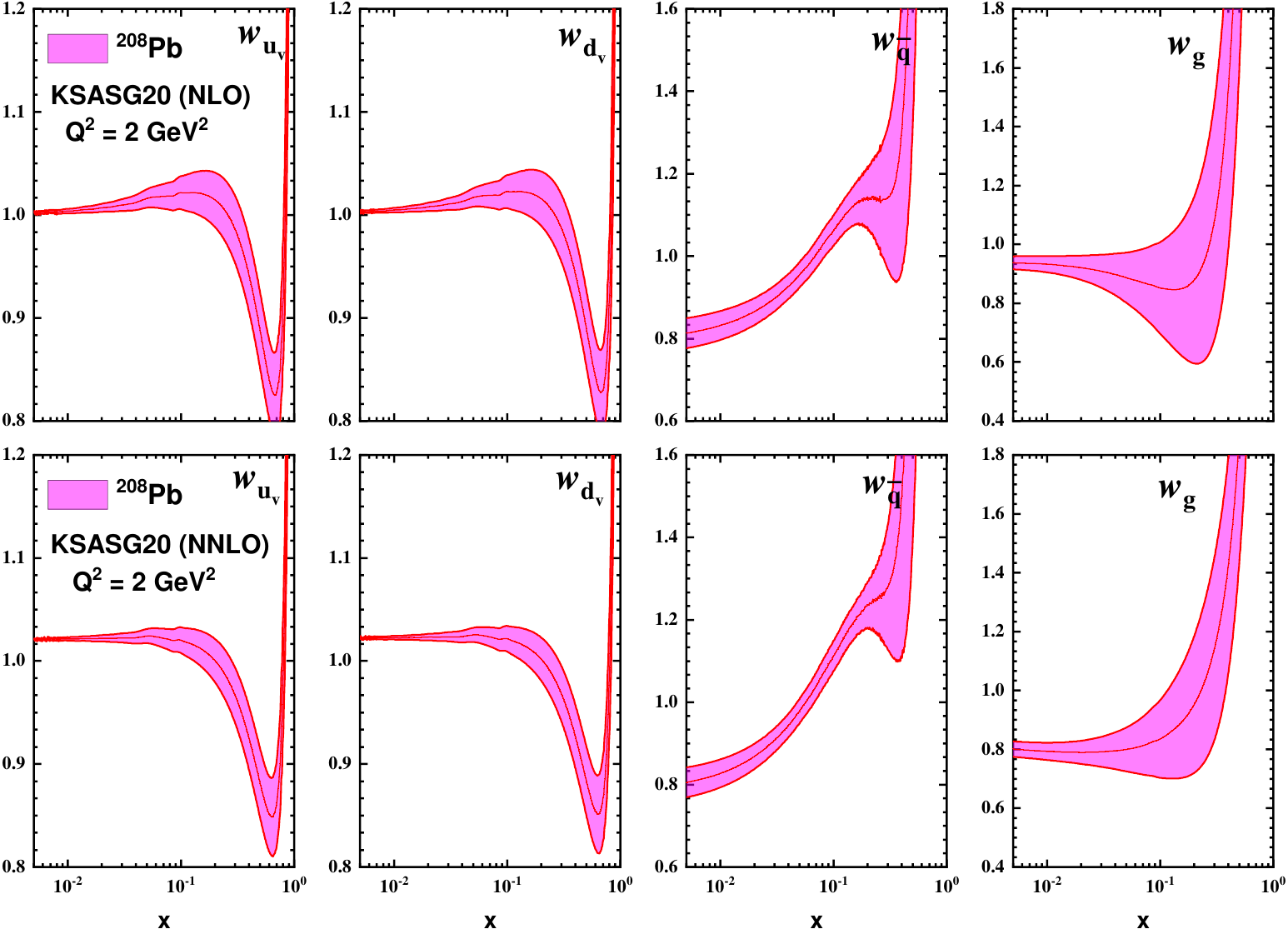}}
	\begin{center}
		\caption{{\small 
			The nuclear modification factors for lead 
			at NLO and NNLO accuracy 
			and at $Q^2$ = 2 GeV$^2$. The bands show the 68\% 
			uncertainty estimation with $\Delta \chi^2 = 20$ obtained 
			using the Hessian method.
			} 
			\label{fig:w-NLO-NNLO-Pb}}  
	\end{center}
\end{figure*}

\begin{figure*}[htb]
\vspace{0.50cm}
\resizebox{0.990\textwidth}{!}{\includegraphics{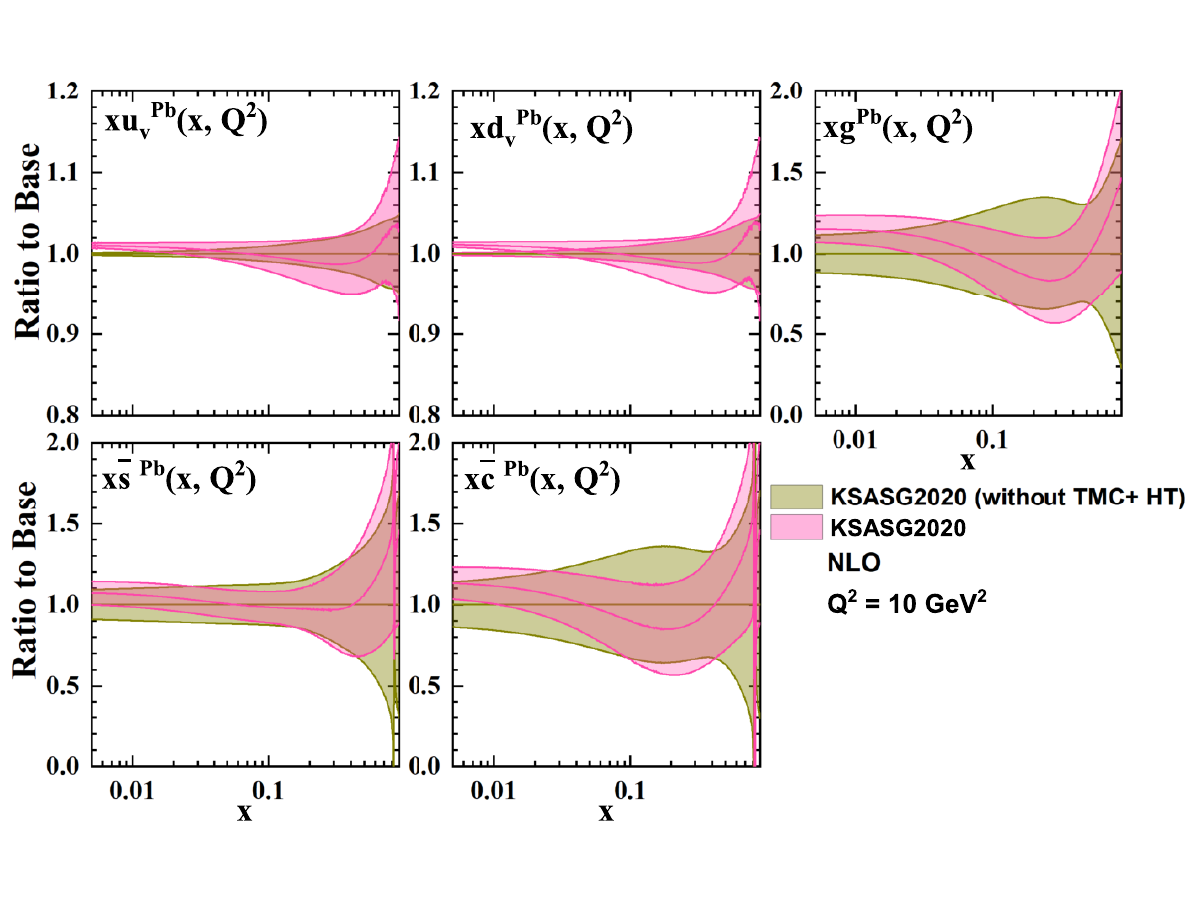}}
\begin{center}
\caption{{\small  
	The ratio of {\tt KSASG20} nuclear PDFs at NLO which include 
	TMC and HT effects compared to the corresponding results 
	without such corrections for lead at $Q^2$ = 10 GeV$^2$.
	We also show the uncertainty bands computed with
	the Hessian method.
	} 
\label{fig:Compare-TMC-HT}}  
\end{center}
\end{figure*}

\begin{figure*}[htb]
\vspace{0.50cm}
\resizebox{0.990\textwidth}{!}{\includegraphics{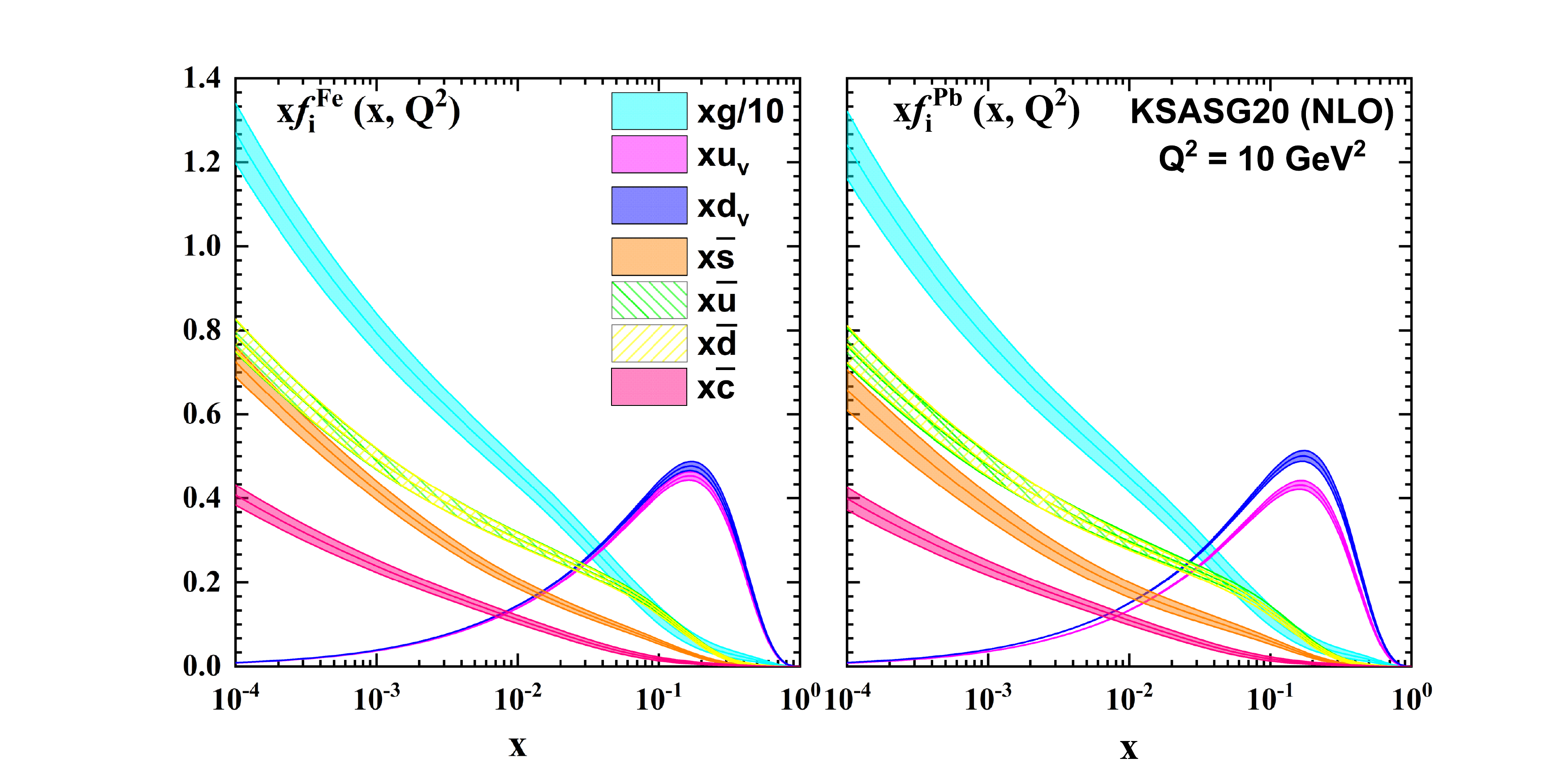}}
\begin{center}
\caption{{\small 
	{\tt KSASG20} nuclear PDFs and their uncertainties at $Q^2$ = 10 
	GeV$^2$ for iron (left) and lead (right). The error bands 
	correspond to the uncertainty estimates at 68\% CL with 
	$\Delta \chi^2 = 20$ obtained using the Hessian method. 
} 
\label{fig:Fe-Pb-Q10}}
\end{center}
\end{figure*}
\begin{figure*}[htb]
\vspace{0.50cm}
\resizebox{0.55\textwidth}{!}{\includegraphics{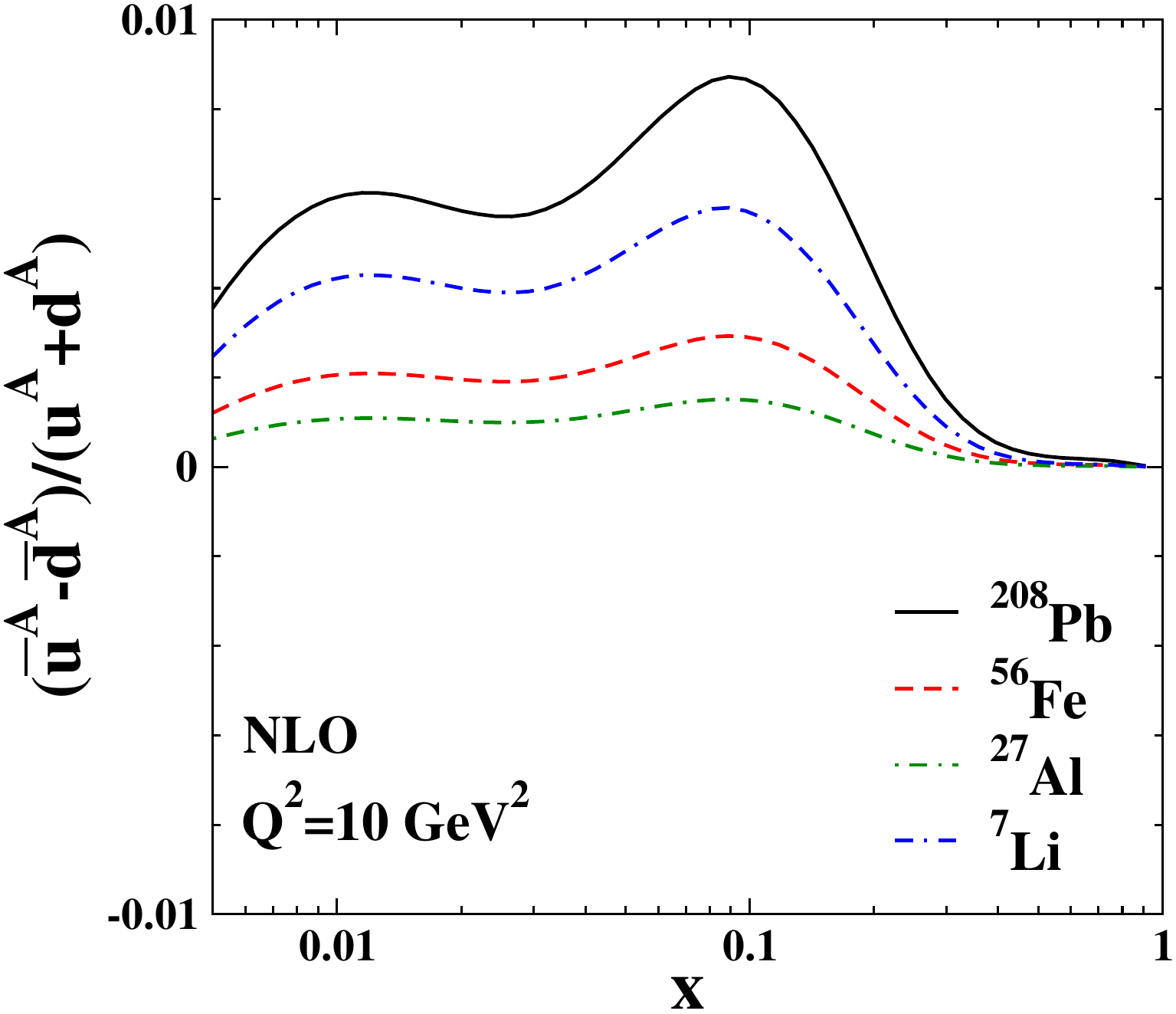}}
\begin{center}
\caption{{\small 
	The flavor asymmetry of sea-quark distributions, 
	$(\bar u^A - \bar d^A)/(u^A + d^A)$, for lead, 
	iron, aluminum and lithium at Q$^2$ = 10~GeV$^2$.
} 
\label{fig:asymmetry}}  
\end{center}
\end{figure*}

\begin{figure*}[htb]
	\vspace{0.50cm}
	\resizebox{0.90\textwidth}{!}{\includegraphics{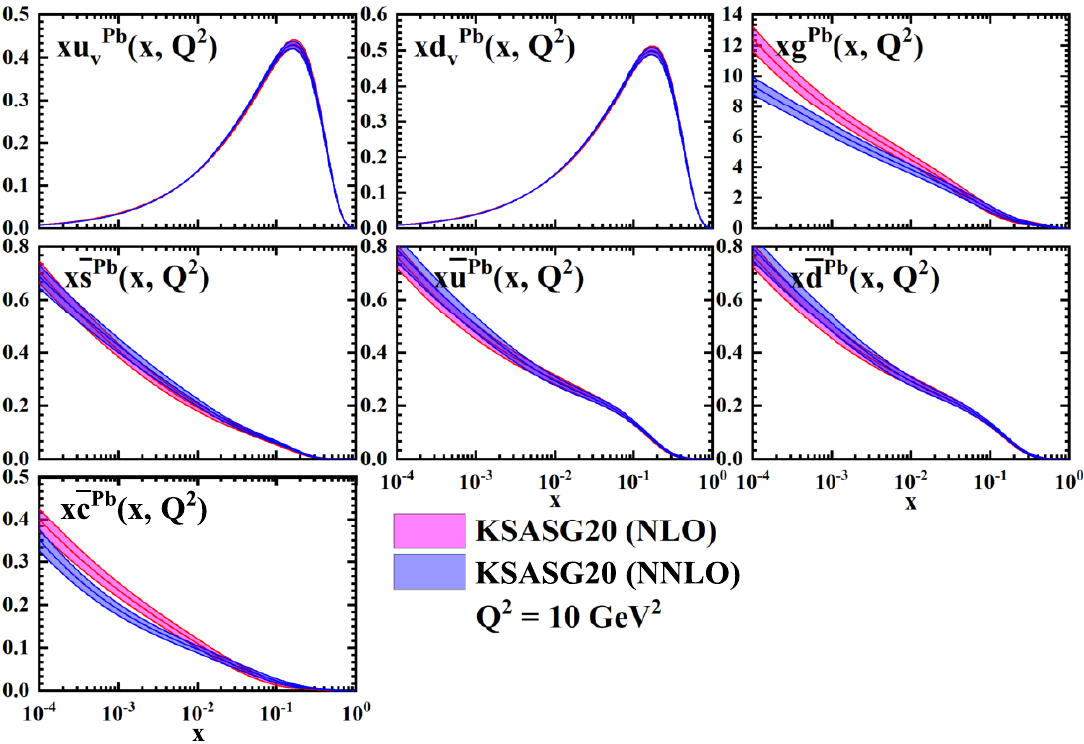}}
	\begin{center}
		\caption{{\small 
			The nuclear PDFs at $Q^2 = 10$~GeV$^2$ for lead at NLO 
			and NNLO accuracy. The uncertainty bands have been 
			calculated with $\Delta \chi^2 = 20$ as described in 
			the text. 
			} 
			\label{fig:parton-nlo-nnlo-Q10-Pb}}
	\end{center}
\end{figure*}

\begin{figure*}[htb]
	\vspace{0.50cm}
	\resizebox{0.90\textwidth}{!}{\includegraphics{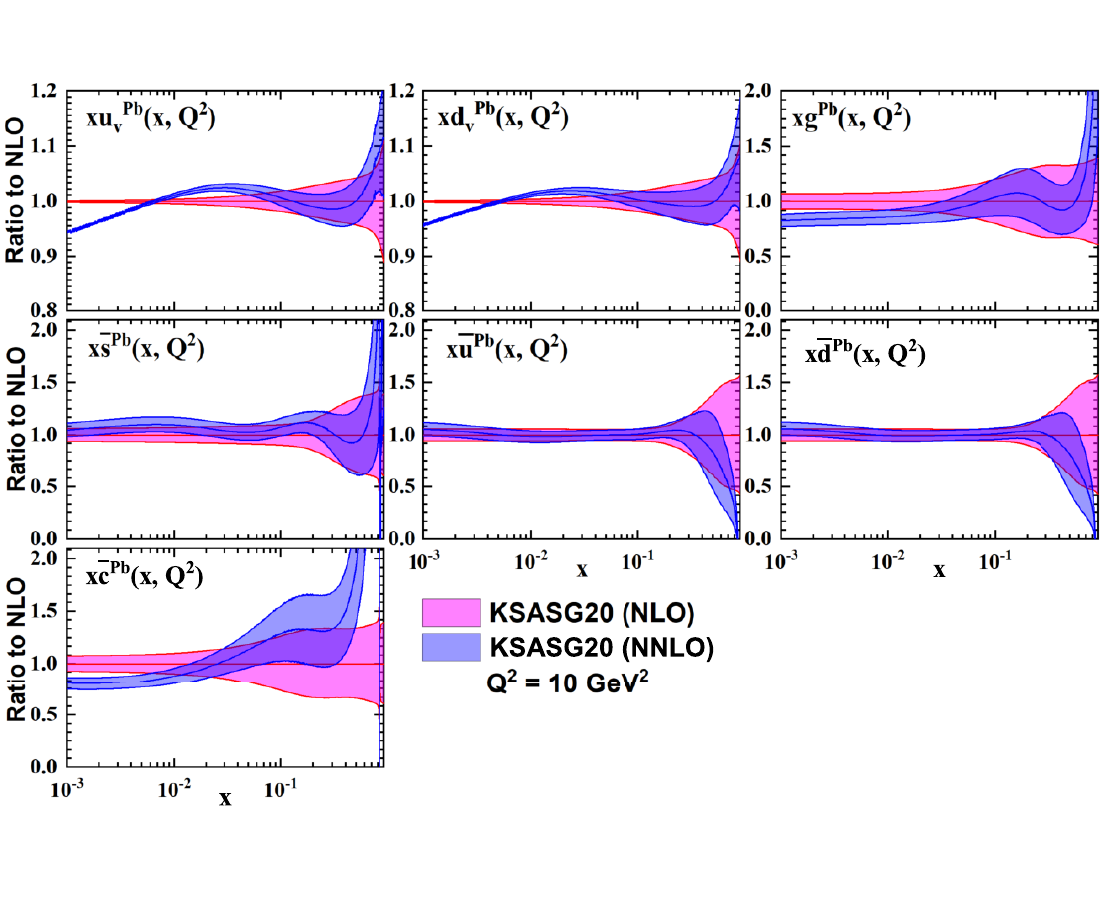}}
	\begin{center}
		\caption{{\small 
			Comparison between the NLO and NNLO nuclear PDFs 
			together with their one-$\sigma$ uncertainties. 
			The results are shown as ratios 
			NNLO/NLO at the scale $Q^2 = 10$ GeV$^2$ for lead. 
			} 
			\label{fig:Ratio-to-NLO-Pb}}
	\end{center}
\end{figure*}

\subsection{ Best fit parameters }\label{sec:fit-parameters}

In this work, we analyze nuclear PDFs using the {\tt CT18} proton 
PDF set as a baseline~\cite{Hou:2019efy}. The nuclear modification 
factors in Eq.~\eqref{eq:weight-function} are extracted from QCD 
fits to the nuclear and neutrino(antineutrino) DIS, 
and Drell-Yan data. Our best 
fit parameters obtained in the {\tt KSASG20} NLO and NNLO fits at 
the initial scale $Q_{0}^{2} = 1.69$ GeV$^2$ are presented in 
Table~\ref{tab-nuclear-pdf} along with their errors. Values marked 
with an asterisk (*) in this table are fixed at the given particular 
value since the analyzed nuclear and neutrino DIS 
and Drell-Yan data could not 
constrain these parameters well enough. The fixed values of 
$\beta_{g} = 1$ and $\beta _{\overline{q}} = 1$ for gluon and sea 
quarks, as well as the value $\beta _{v} = 0.81$ for valence 
quark densities is  motivated by the {\tt HKN07} 
analysis~\cite{Hirai:2007sx}. 
Freeing these parameters can easily 
lead to unphysical fit results, and hence, we have decided to keep 
them fixed at this stage. As we mentioned before, the heavy-quark 
masses are fixed at $m_c=1.30$ GeV and $m_b=4.75$ GeV to be 
consistent with the {\tt CT18} proton PDFs. The strong coupling 
constant is taken as 
$\alpha_s (M_Z)=0.118$~\cite{Tanabashi:2018oca}.   

As discussed in Sec.~\ref{sec:PDFs-nucleus}, the nucleus-dependent 
parameters $a_i(A, Z)$ for the sea quark densities need to be 
determined from the fit to data. The parameters $a_i(A, Z)$ for 
the valence quark and gluon densities depend on the mass number $A$ 
and atomic number $Z$ in general and are extracted from the three 
constraints in Eqs.~\eqref{eq:nuclear-charge}, 
\eqref{eq:Baryon_number} and \eqref{eq:momentum-sum-rule}.
The numerical values for these parameters are presented in 
Tables~\ref{tab:au-ad-ag-NLO} and \ref{tab:au-ad-ag-NNLO} at NLO 
and NNLO accuracy, respectively. 

Regarding the best fit parameters and their errors listed in 
Table~\ref{tab-nuclear-pdf}, some comments are in order. The 
obtained parameters for the nuclear valence-quark distributions 
reflect the fact that the nuclear DIS data analyzed in this study 
constrain these distributions well enough. In addition, 
neutrino DIS  and Drell-Yan  data also play
an important role in obtaining a consistent 
behaviour for the up and down valence quarks.  As can be seen from 
Table~\ref{tab-nuclear-pdf}, some fit parameters of our sea-quark 
and gluon densities come with larger errors, especially $\epsilon_{d_{\overline{q}}}$  and 	$\epsilon_{c_{g}}$.
In addition, we fixed $d_g$ to zero. 
The nuclear DIS and Drell-Yan data 
only loosely constrain the gluon nuclear modifications because they 
cover a too limited range in $Q^2$. 

To further constrain the nuclear sea-quark and gluon densities 
and reduce their uncertainties, other observables will have to 
be taken into account. 
Once more data are included, for example the data from the LHC 
and a future $eA$ collider such as LHeC or FCC-he, it should be 
possible to relax some of the assumptions mentioned above.

\begin{figure*}[htb]
\vspace{0.50cm}
\resizebox{0.90\textwidth}{!}{\includegraphics{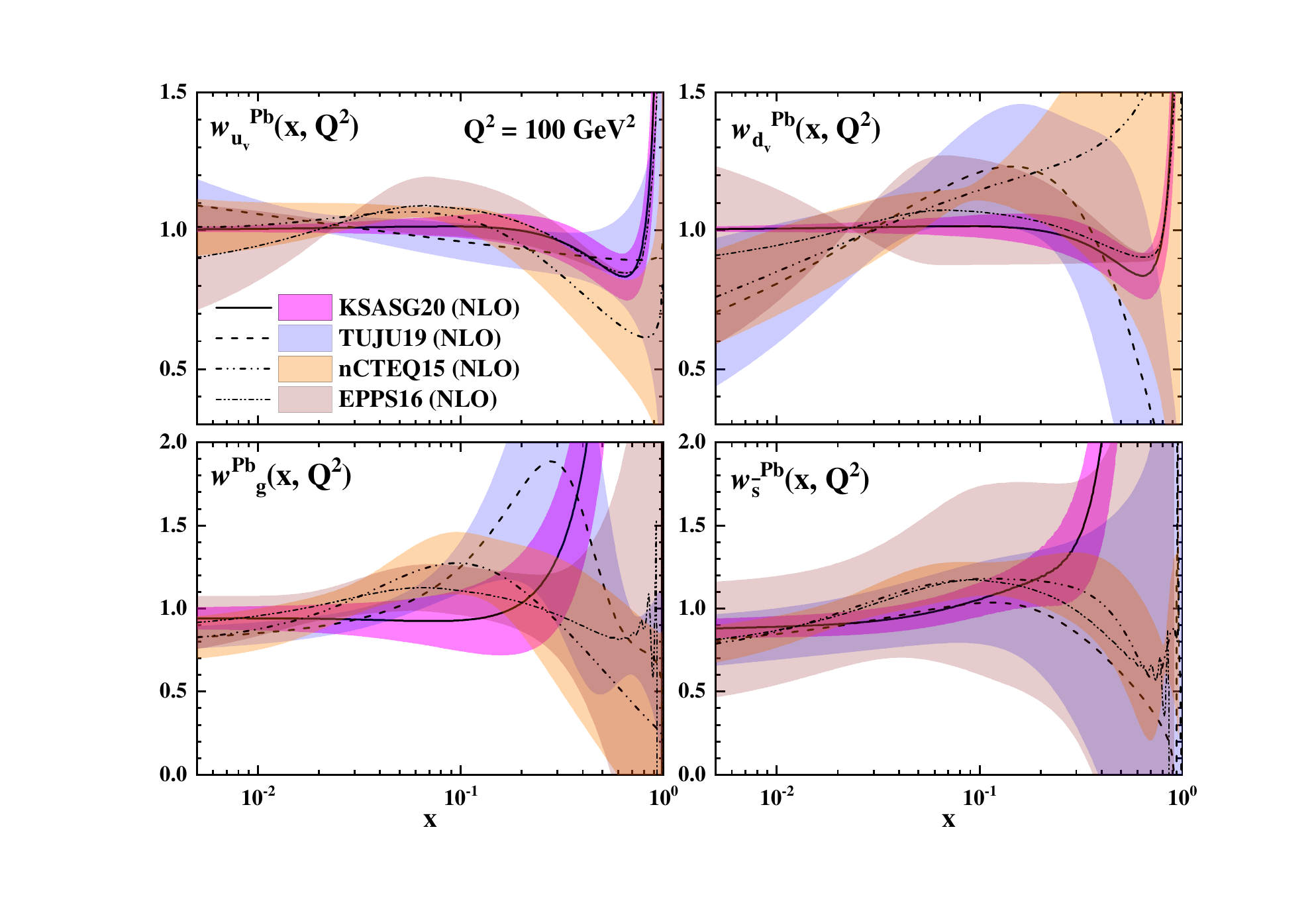}}
\begin{center}
\caption{{\small  
	The nuclear modification factors {\tt KSASG20} in 
	lead at NLO accuracy compared to the nuclear PDF sets 
	{\tt nCTEQ15}~\cite{Kovarik:2015cma},   
	{\tt TUJU19}~\cite{Walt:2019slu} and {\tt EPPS16}~\cite{Eskola:2016oht} 
	shown at the scale 
	$Q^{2} = 100~{\rm GeV}^{2}$. The comparison is presented 
	per parton flavor $i$.
	The bands for the 
	{\tt KSASG20} show the 68\% uncertainty estimation with 
	$\Delta \chi^2 = 20$ obtained using the Hessian method. 
} 
\label{fig:Modification-Model-TUJU-nCTEQ-EPPS-NLO}}  
\end{center}
\end{figure*}

\begin{figure*}[htb]
	\vspace{0.50cm}
	\resizebox{0.90\textwidth}{!}{\includegraphics{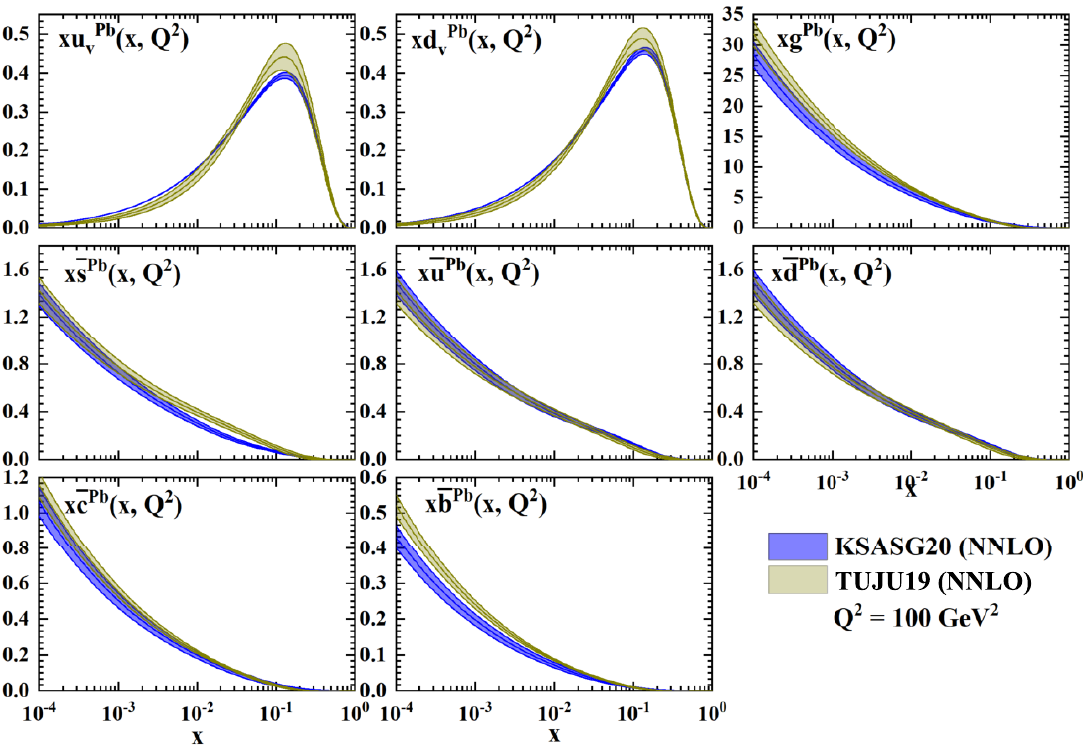}}
	\begin{center}
		\caption{{\small 
			The nuclear PDFs at the scale $Q^2 = 100$~GeV$^2$ 
			for lead at NNLO accuracy, compared with the recent 
			results from {\tt TUJU19}~\cite{Walt:2019slu}. 
			} 
			\label{fig:Compare-Model-TUJU-NNLO}}
	\end{center}
\end{figure*}
\begin{figure*}[htb]
	\vspace{0.50cm}
	\resizebox{0.990\textwidth}{!}{\includegraphics{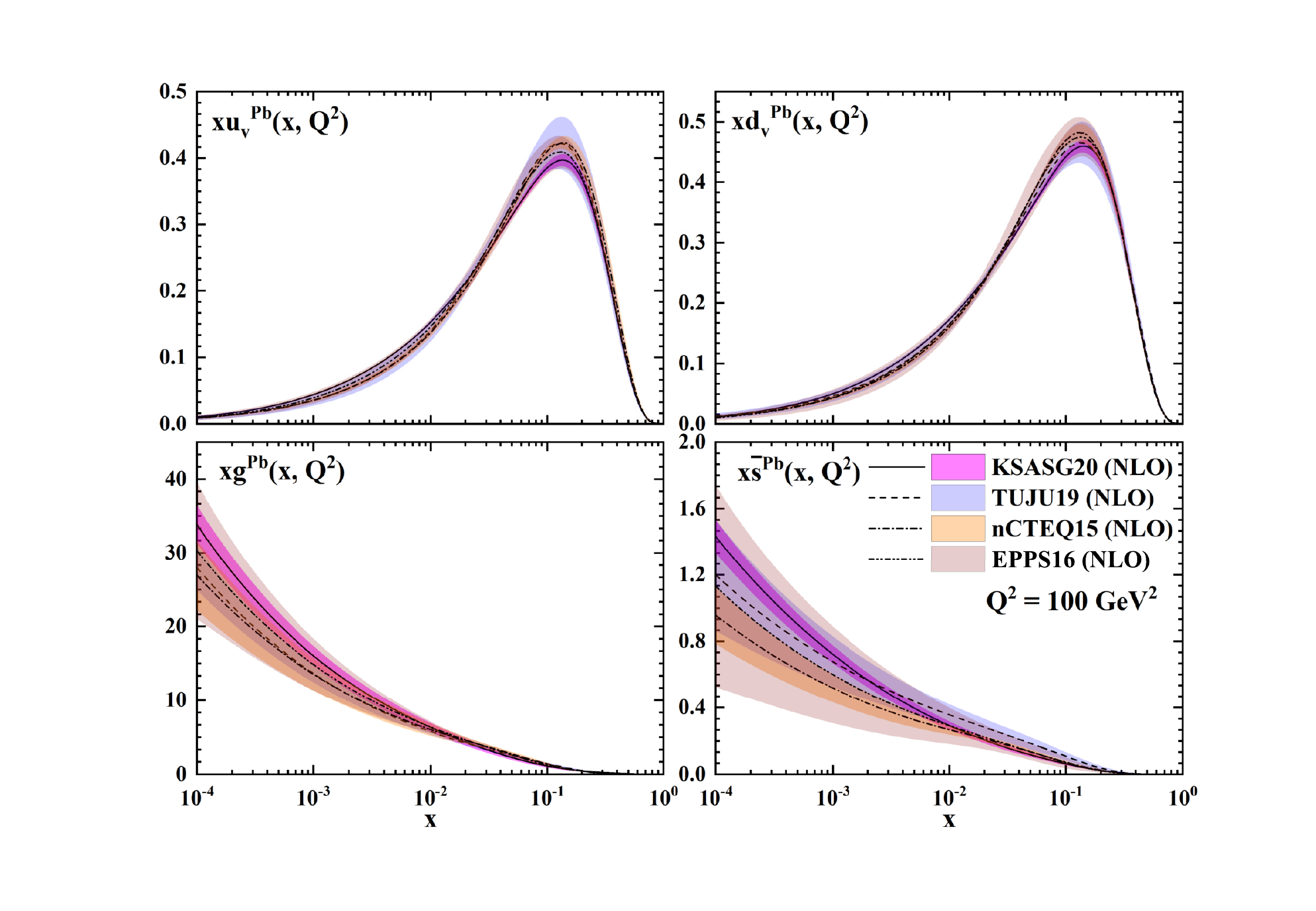}}
	\begin{center}
		\caption{{\small 
			The nuclear PDFs at the scale $Q^2 = 100$~GeV$^2$ 
			for lead at NLO accuracy, compared with results from 
			{\tt nCTEQ15}~\cite{Kovarik:2015cma}, 
			{\tt EPPS16}~\cite{Eskola:2016oht} and 
			{\tt TUJU19}~\cite{Walt:2019slu}. 
			The comparison is presented for the parton flavors 
			$u_v$, $d_v$, $g$ and $\bar{s}$. 
			} 
			\label{fig:Compare-Model-TUJU-nCTEQ-EPPS-NLO-1}}
	\end{center}
\end{figure*}
\begin{figure*}[htb]
	\vspace{0.50cm}
	\resizebox{0.990\textwidth}{!}{\includegraphics{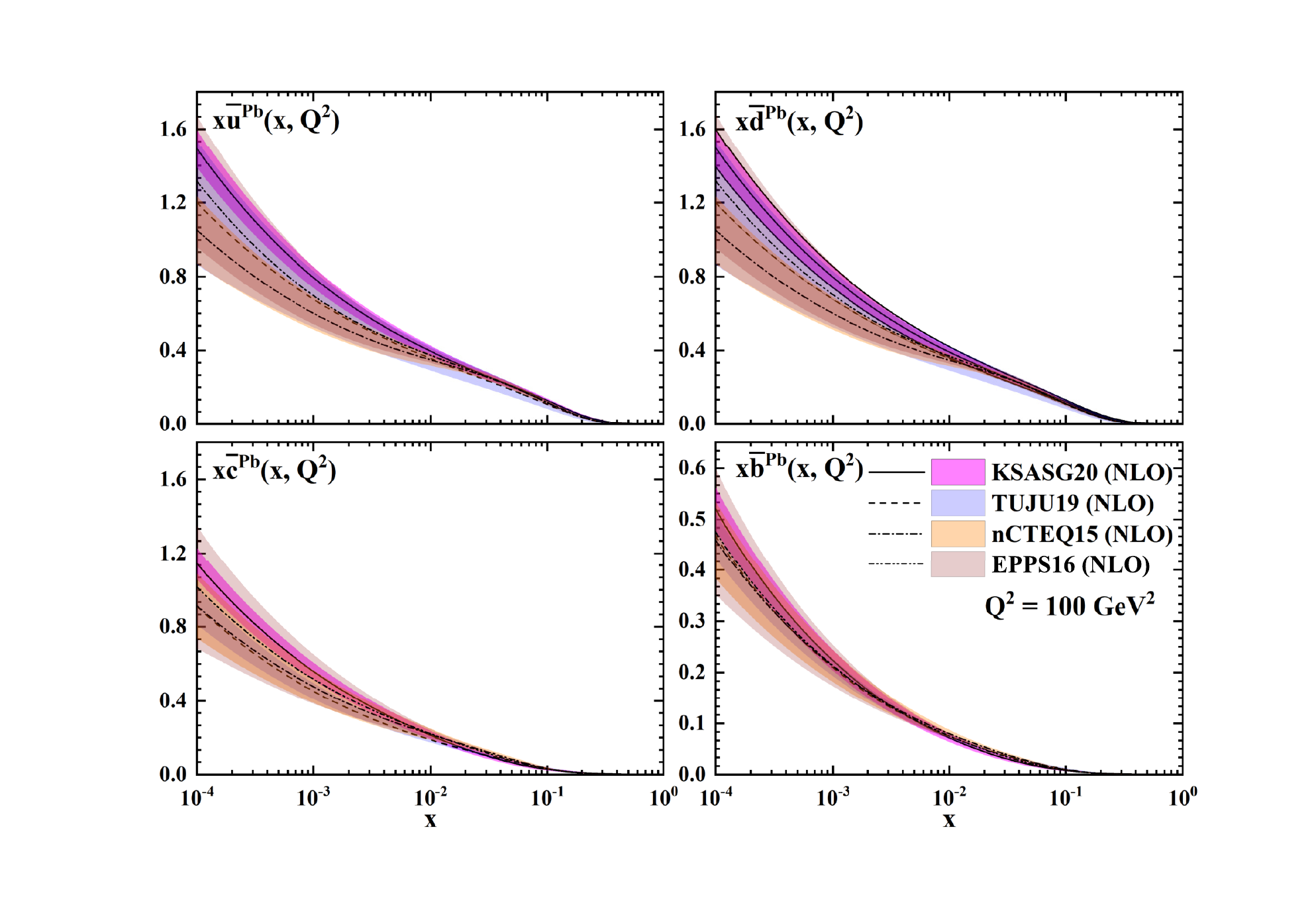}}
	\begin{center}
		\caption{{\small  
				Same as Fig.~\ref{fig:Compare-Model-TUJU-nCTEQ-EPPS-NLO-1}, but 
				for the $\bar{u}$, $\bar{d}$, $\bar{c}$, and $\bar{b}$ PDFs. 
			}
			\label{fig:Compare-Model-TUJU-nCTEQ-EPPS-NLO-2}}
	\end{center}
\end{figure*}
\begin{figure*}[htb]
	\vspace{0.50cm}
	\resizebox{0.990\textwidth}{!}{\includegraphics{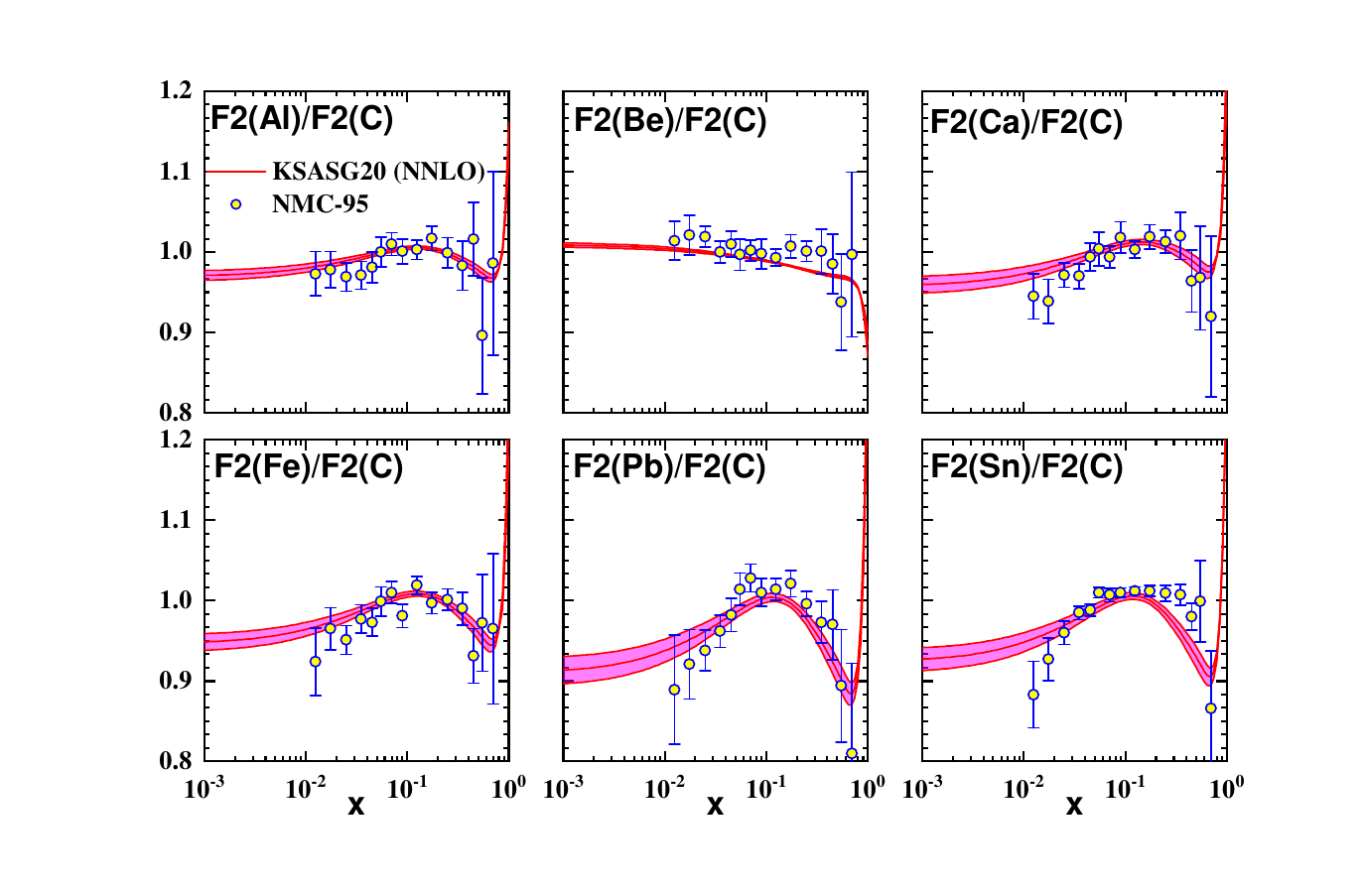}}
	\begin{center}
		\caption{{\small 
			Comparison of our NNLO results for the ratio 
			$F_2^A(x, Q^2) / F_2^{C} (x, Q^2)$ as a function of 
			$x$ with some selected nuclear DIS data from NMC-95. 
			Our NNLO results  have 
			been calculated at Q$^2$=5 GeV$^2$. 
			The bands show the 68\% uncertainty estimate with 
			$\Delta \chi^2 = 20$. 
			} 
			\label{fig:Ratio-F2A-over-F2C}}
	\end{center}
\end{figure*}
\begin{figure*}[htb]
	\vspace{0.50cm}
	\resizebox{0.990\textwidth}{!}{\includegraphics{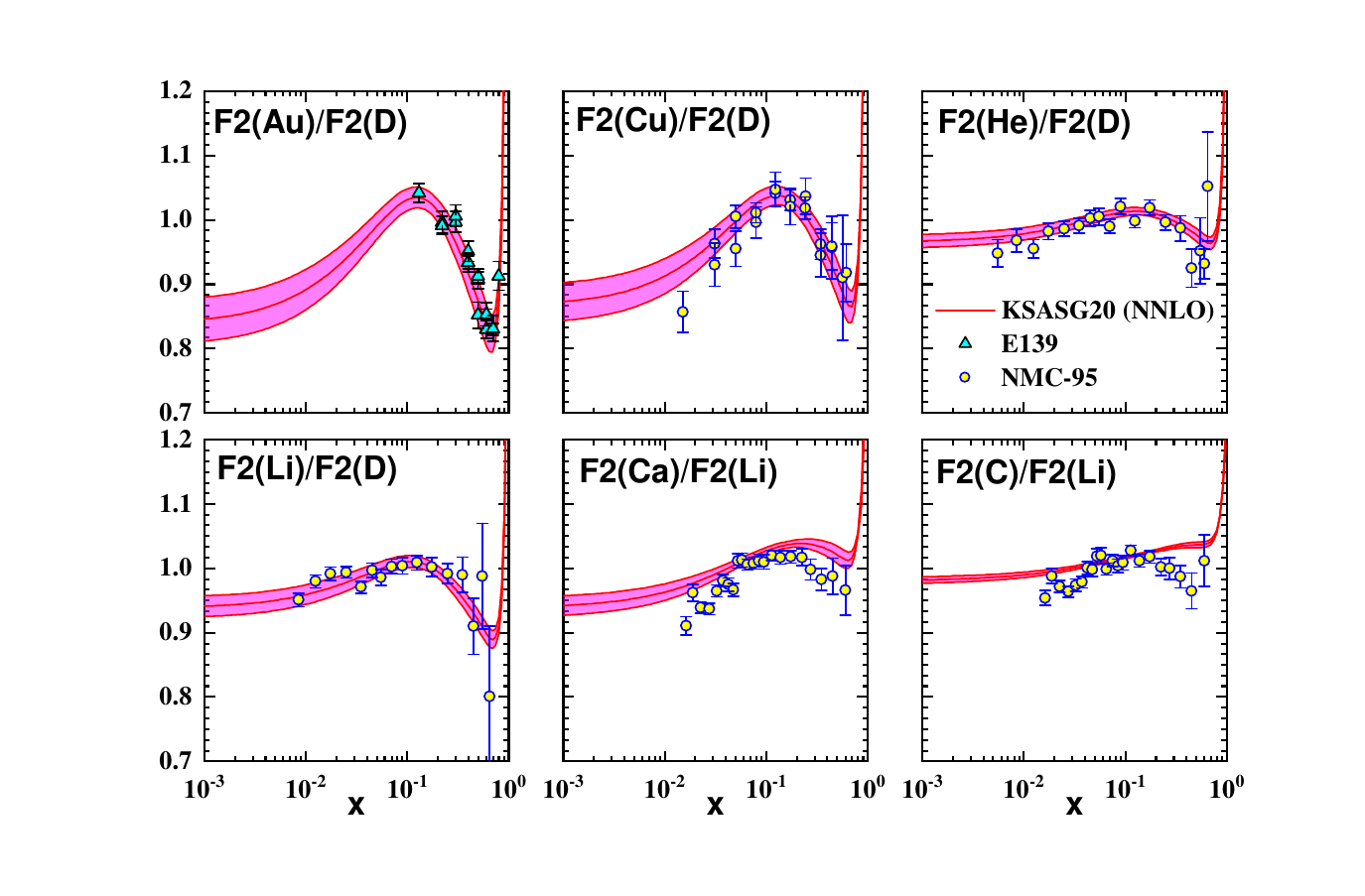}}
	\begin{center}
		\caption{{\small 
			Comparison of our NNLO results  
			calculated at $Q^2$ = 5 GeV$^2$ for the ratios 
			$F_2^A(x, Q^2) / F_2^{D} (x, Q^2)$ and $F_2^A(x, Q^2) / 
			F_2^{Li} (x, Q^2)$ as a function of $x$ with some 
			selected nuclear DIS data from E139 and NMC-95. 
			} 
			\label{fig:Ratio-F2A-over-F2D}}
	\end{center}
\end{figure*}

\begin{figure*}[htb]
	\vspace{0.50cm}
	\resizebox{0.90\textwidth}{!}{\includegraphics{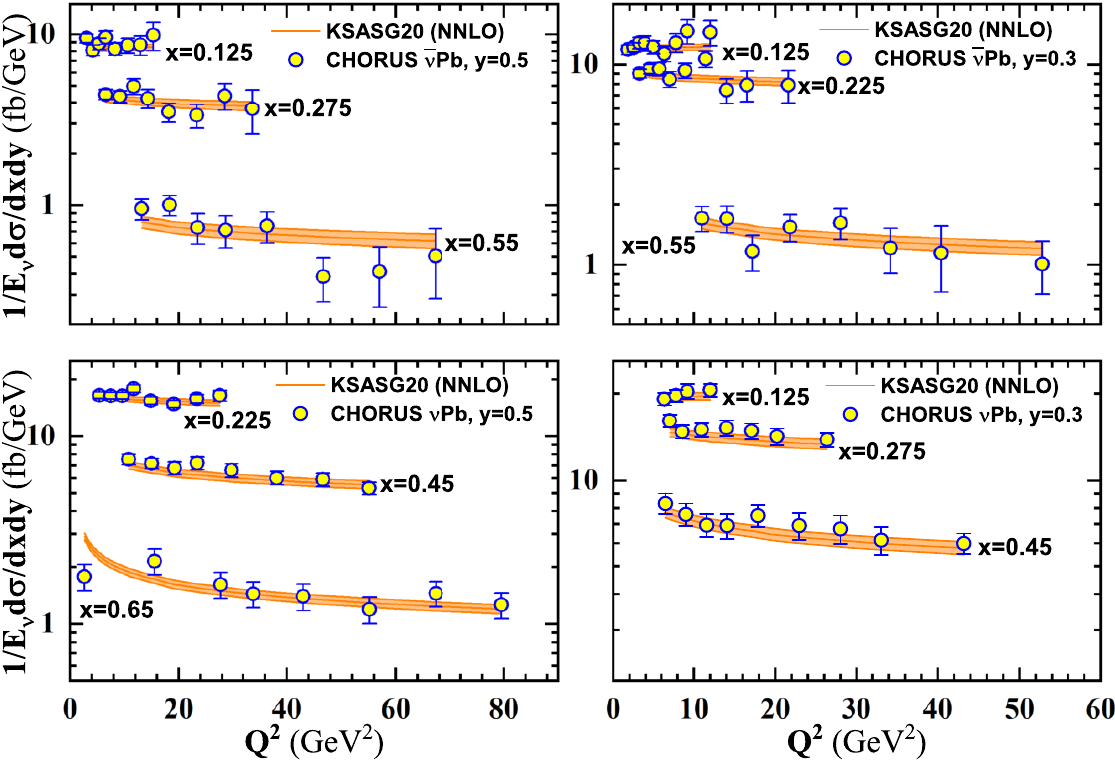}}
	\begin{center}
		\caption{{\small 
				The neutrino(antineutrino) lead DIS data from the CHORUS 
				measurements~\cite{Onengut:2005kv}, compared with the 
				{\tt KSASG20} NNLO results. The bands show the 
				68\% uncertainty estimates with $\Delta \chi^2 = 20$. 
			} 
			\label{fig:Neutrino-Antineutrino-Lead}}
	\end{center}
\end{figure*}
\begin{figure*}[htb]
	\vspace{0.50cm}
	\resizebox{0.90\textwidth}{!}{\includegraphics{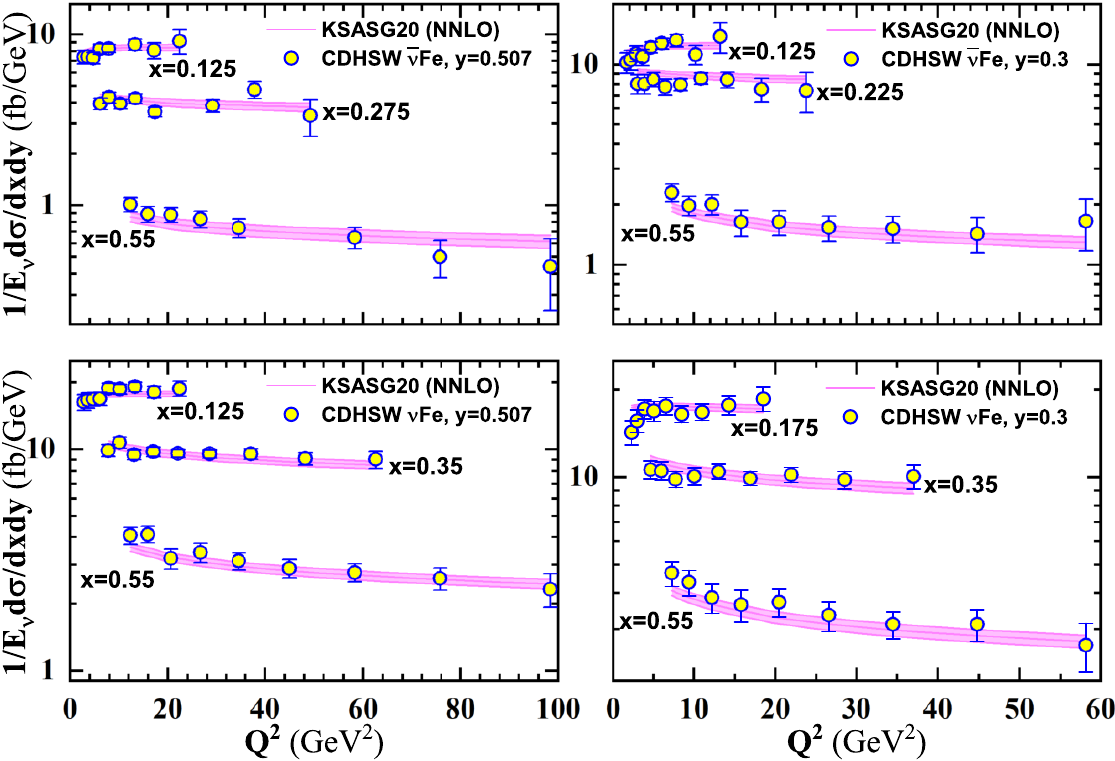}}
	\begin{center}
		\caption{{\small 
				Same as Fig.~\ref{fig:Neutrino-Antineutrino-Lead}, but for the 
				neutrino(antineutrino) iron DIS data from the CDHSW  
				measurements~\cite{Berge:1989hr}. 
			} 
			\label{fig:Neutrino-Antineutrino-Iron}}
	\end{center}
\end{figure*}

\begin{figure*}[htb]
\vspace{0.50cm}
\resizebox{0.90\textwidth}{!}{\includegraphics{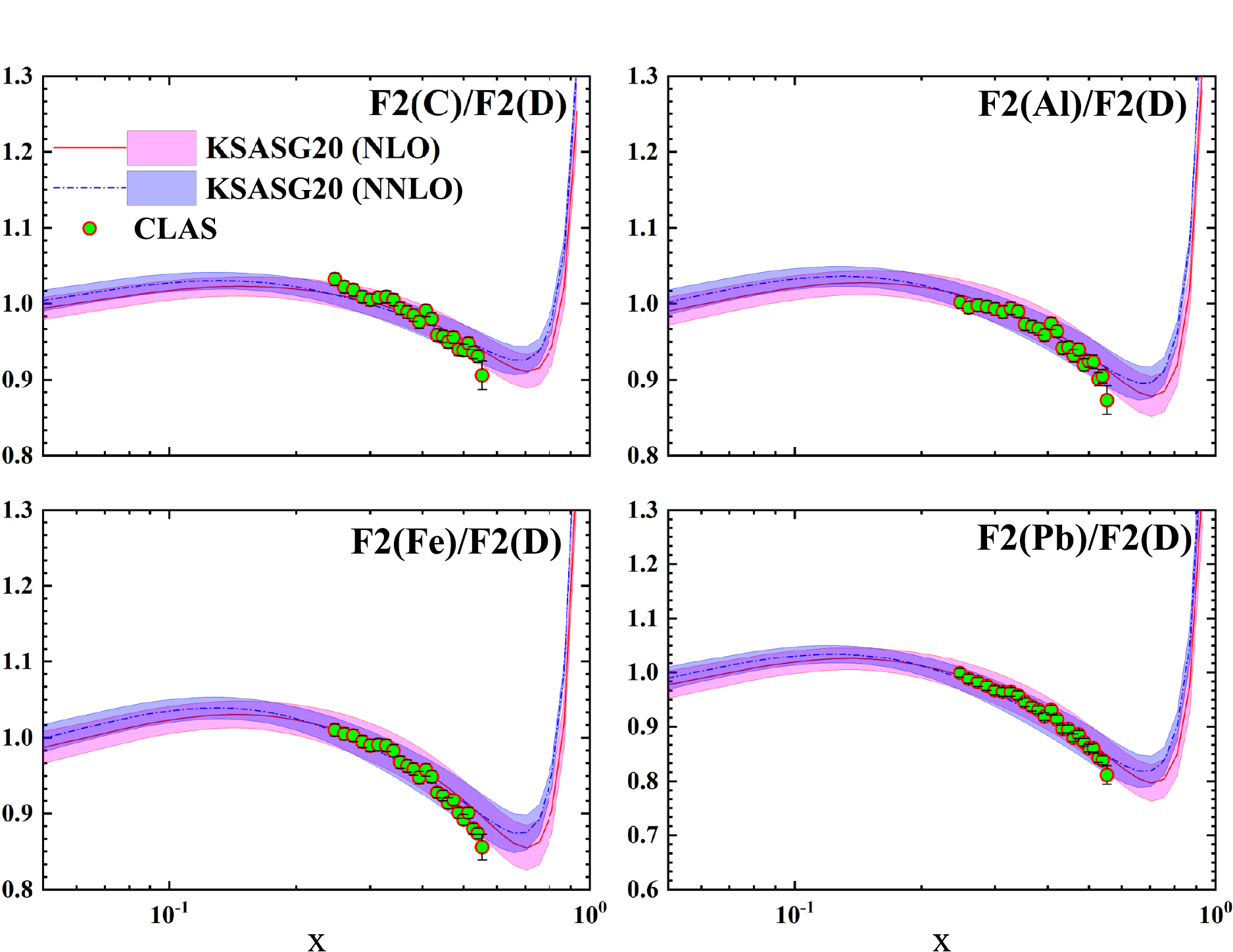}}
\begin{center}
\caption{{\small 
	Comparison of our NLO and NNLO results  
	calculated at $Q^2$ = 3 GeV$^2$ for the structure function 
	ratios $F_2^A(x, Q^2) / F_2^{D} (x, Q^2)$ with the data from 
	Jefferson Lab CLAS~\cite{Schmookler:2019nvf} as a function 
	of $x$. 
} 
\label{fig:F2-A-F2-D-CLAS-Final}}
\end{center}
\end{figure*}

\begin{figure*}[htb]
\vspace{0.50cm}
\resizebox{1.0\textwidth}{!}{\includegraphics{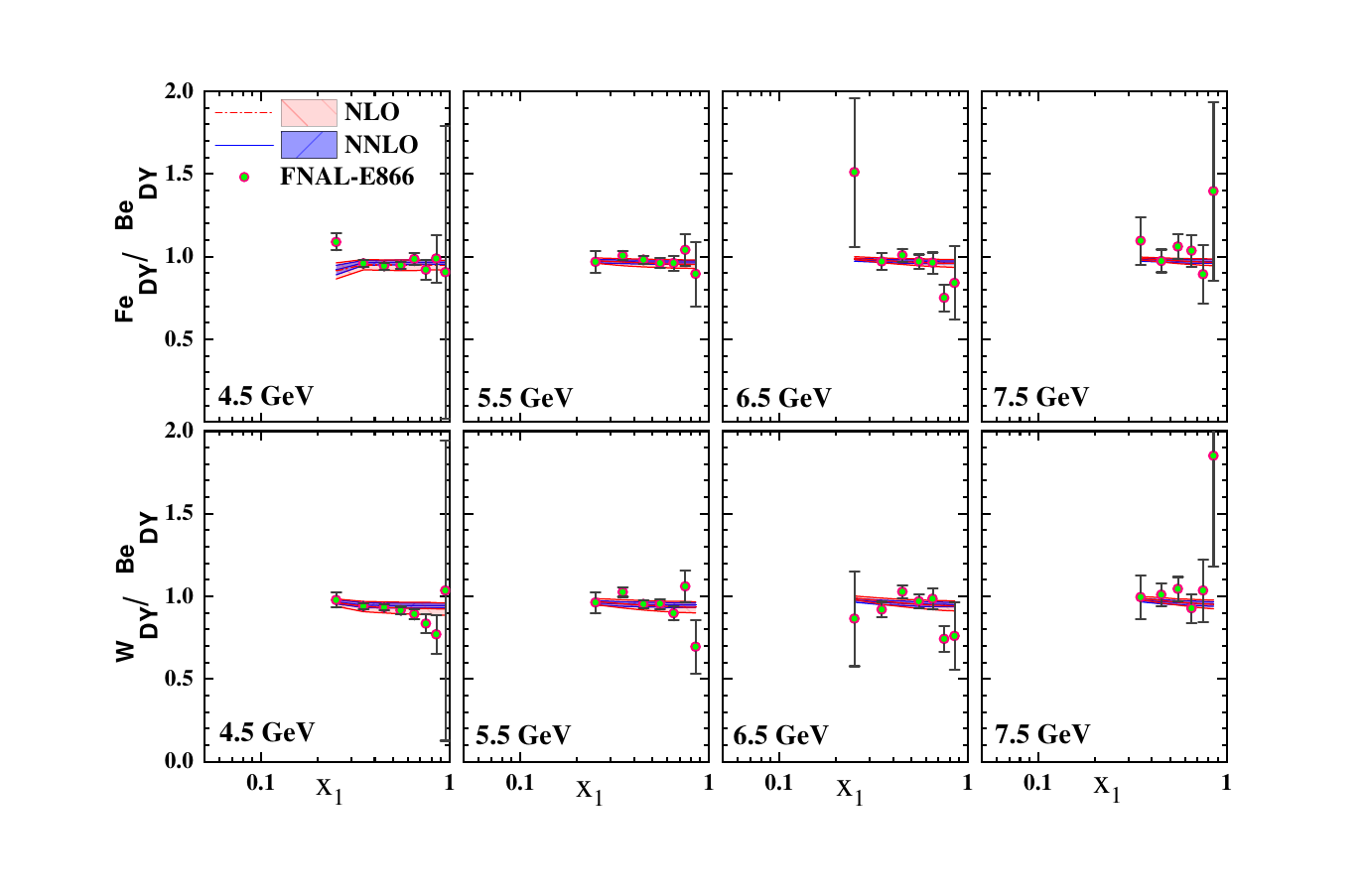}}
\begin{center}
\caption{{\small  
	Comparison of the {\tt KSASG20} NLO and NNLO results 
	along with their uncertainties for the 
	Drell-Yan cross section ratios 
	$\sigma_{{\text{DY}}}^A/\sigma_{{\text{DY}}}^{A^{\prime}}$ 
	with data 
	for some selected nuclear targets from the Fermilab experiment 
	E866~\cite{Vasilev:1999fa}. 
} 
\label{fig:DY}}
\end{center}
\end{figure*}

\subsection{ {\tt KSASG20} nuclear PDFs and their uncertainties }\label{sec:KSASG20-nuclear-PDFs}

In the following, we discuss the {\tt KSASG20} nuclear PDFs 
including the nuclear modification functions and present a 
detailed comparison between our NLO and NNLO analyses. 

In Fig.~\ref{fig:w-NLO-NNLO}, we show representations  
of different 
types of nuclear modification functions for some selected nuclei, 
deuterium (D), beryllium (Be), iron (Fe) and gold (Au) at the scale 
$Q^2$ = 2 GeV$^2$. The nuclear modification functions are shown for 
the valence-quark ${\cal W}_{u_v}$ and ${\cal W}_{d_v}$, sea-quark 
${\cal W}_{\bar q}$ and gluon ${\cal W}_{g}$ at NLO (top row) and 
NNLO (bottom row). We repeat here that, in this work, we 
have treated deuterium as a nucleus in the fitting procedure. 
Hence, as one can see from Fig.~\ref{fig:w-NLO-NNLO}, small 
deviations from the baseline {\tt CT18} proton PDFs are found for 
deuterium. The deviations from the {\tt CT18} PDFs become 
larger with increasing atomic mass, and significant effects 
are found for heavier nuclei, such as gold.
We should notice here that our results in the small-$x$ region,  
i.e.\ $x < 10^{-2}$, are not directly constrained by the nuclear 
and neutrino DIS and Drell-Yan data, 
but determined by extrapolation based on 
our parametrization.  

As can be seen from Fig.~\ref{fig:w-NLO-NNLO}, 
the typical nuclear modification effects, such as anti-shadowing, 
shadowing and EMC suppression, are visible  at the 
$Q^2 = 2$ GeV$^2$ for the up and down valence quark modifications, 
${\cal W}_{u_v}$ and ${\cal W}_{d_v}$. For the NLO analysis, the 
nuclear modification for the gluon shows a rapid rise with increasing 
$x$, $x > 0.1$. This trend repeats itself for the NNLO analysis, 
a behaviour which is similar to what one can observe in the analyses 
by {\tt HKM01}~\cite{Hirai:2001np}, {\tt HKN07}~\cite{Hirai:2007sx} 
and {\tt KA15}~\cite{Khanpour:2016pph}. This behaviour may be an 
artefact due to the used parametrization in these analyses. 
Figure~\ref{fig:w-NLO-NNLO} also shows that for the case of sea-quarks 
one can observe the typical nuclear modifications. However, the 
magnitude of these effects slightly differs at different 
perturbative orders. 
Our nuclear modifications for quark and gluon 
densities are flat in the small-$x$ region. This behaviour is 
similar to the analysis of {\tt HKN07}~\cite{Hirai:2007sx}.

We continue with the discussion of the nuclear modification factors 
and their uncertainties for the case of lead (Pb) 
as an example of a large nucleus. 
Lead is particularly relevant for the present and future 
heavy-ion program at the LHC for p-Pb and Pb-Pb collisions. 
The nuclear modification factors for lead 
along with their uncertainties at 68\% CL with $\Delta \chi^2 = 20$ 
are shown in Fig.~\ref{fig:w-NLO-NNLO-Pb}. The results are shown 
at the scale $Q^2$ = 2 GeV$^2$ at NLO and NNLO accuracy. 
The uncertainty bands at very small and large $x$ are not 
directly constrained by data. They are affected by the 
restricted flexibility of the considered parametrization 
and the limitations of the fitting framework.
Due to the limited sensitivity of the available nuclear data to the 
sea-quark and gluon densities, one has to limit the number
of shape parameters. 
One can expect that additional new data will provide better 
constraints and one can envisage to consider more fit parameters
allowing a larger flexibility of the modification functions.

In the following we discuss the effect of TMC and HT corrections 
on the fit quality and the shape of the extracted nuclear PDFs. 
At NLO, we observe that the inclusion of TMCs slightly 
increases $\chi^2/N_{dof}$ from 1.02 to 1.06. This can be traced 
back to the JLab  Hall C and CLAS data. Including the HT 
corrections of Sec.~\ref{sec:Higher-Twist-Corrections}
yields a small improvement: $\chi^2/N_{dof}$ is reduced from 
1.06 to 1.05 and the TMC and TMC+HT fits are quite similar. 
The same behaviour is seen in the case of the NNLO fit. 
We conclude that the inclusion of TM and HT corrections is 
not crucial for a good description of the presently available 
high-$x$ data \cite{Paukkunen:2020rnb}. 

The effect of the inclusion of TMC and HT on the 
extracted nuclear PDFs is apparent from
Fig.~\ref{fig:Compare-TMC-HT}, where we compare our fit without TMC and HT 
({\tt KSASG20 without TMC+HT}) with our final fit, 
which includes both of these corrections ({\tt KSASG20}). 
The first observation emerging from the comparisons presented in this plot
is that while the TMC and HT affect the large $x$ region,  
they do not significantly change the results at small $x$.
As one can see, the {\tt KSASG20} valence quark PDF is slightly larger than 
({\tt KSASG20 without TMC+HT}) 
and this effect is mostly localized at large values of $x$ as expected.
Also the gluon and the charm-quark PDFs are modified by TMC+HT 
effects at large, while somewhat suppressed at intermediate 
values of $x$.

The resulting nuclear PDFs are presented in Fig.~\ref{fig:Fe-Pb-Q10} 
for iron (left) and lead (right) at $Q^2$ = 10 GeV$^2$ to 
show the effects of the $Q^2$ evolution.
As we mentioned in Sec.~\ref{sec:PDFs-nucleus}, the 
up and down nuclear PDFs have been assumed to be flavor dependent, 
{\it i.e.}, $x \bar{d} \neq x \bar{u}$. For the strange quark 
distributions in the nuclei, we assume as usual $x s = x \bar{s}$. 
The perturbatively generated heavy quark densities, $x \bar{c}$ and 
$x\bar{b}$, are obtained through DGLAP evolution. 
As can be seen in the figure, all the gluon and sea-quark densities 
come with relatively large error bands at small $x$, reflecting 
the fact that there are not enough data constraints below 
$x \sim 0.01$. We find only very small differences for the 
fitted $\bar{u}$ and $\bar{d}$ nuclear PDFs; the corresponding 
error bands shown in Fig.~\ref{fig:Fe-Pb-Q10} are difficult to 
distinguish. The small differences between $\bar{u}$ and $\bar{d}$ 
nuclear PDFs mainly come from the underlying free proton PDFs and 
the different number of protons and neutrons in  different 
nuclei. From our definition, the $\bar{u}$ and  $\bar{d}$ PDFs are 
equal for isoscalar nuclei such as calcium and carbon.
To show the flavor asymmetry, in Fig.~\ref{fig:asymmetry}, 
we present the ratio $(\bar u^A - \bar d^A)/(u^A + d^A)$ for 
lead (Pb), iron (Fe), aluminum (Al) and lithium (Li) at 
$Q^2 = 10$~GeV$^2$. 

In the following, we compare our NLO and NNLO results.
In Fig.~\ref{fig:parton-nlo-nnlo-Q10-Pb}, the NLO and NNLO 
nuclear PDFs are compared at $Q^2=10$ GeV$^2$ for the valence 
quark $xq_v$, gluon $xg$, strange quark $x \bar{s}$, sea quarks 
$x\bar{u}$ and $x\bar{d}$, and finally the perturbatively generated 
charm quark $x \bar{c}$ density. It is worth noticing here that 
the magnitude and the shape of the nuclear PDFs for a given flavor 
at some arbitrary scale $Q^2$ depends on the chosen set of 
reference PDFs for the free proton. 
Due to the limitations of the applied fitting
framework and the limited sensitivity 
of the analyzed nuclear and neutrino(antineutrino) DIS, and 
Drell-Yan data to the gluon and sea-quark nuclear PDFs, the 
provided uncertainty bands are rather large, especially for 
small values of $x$.

A few remarks concerning the comparison between our NLO and NNLO 
analyses are in order. For both lead and gold nuclei, the valence 
quark, $xu_v$ and $xd_v$, and strange quark $x \bar{s}$ PDF 
densities at NLO and NNLO accuracy are very similar in size. The 
sea quark densities, $x \bar{u}$ and $x \bar{d}$, are slightly 
different at NLO and NNLO accuracy in the region of small $x$, 
$x < 0.01$.  A significant difference can be found for the gluon 
$xg$ and the perturbatively generated charm quark $x \bar{c}$ 
density at NLO and NNLO accuracy. As can be seen from 
Fig.~\ref{fig:parton-nlo-nnlo-Q10-Pb}, the NNLO gluon and charm quark 
PDFs are smaller than at NLO at small values of $x$.

To quantify the magnitude of NNLO corrections,
we also present ratios of nuclear PDFs obtained in the NNLO fit 
over those of the NLO fit. This comparison is shown in 
Fig.~\ref{fig:Ratio-to-NLO-Pb} for lead. 
The results are displayed 
at the scale $Q^2 = 10$ GeV$^2$ and we include the one-$\sigma$ 
uncertainty bands for $\Delta \chi^2 = 20$. 
As can be seen, the 
uncertainty for the nuclear gluon density slightly decreases when 
going from NLO to NNLO accuracy
due to the improved
overall fit quality when higher-order QCD calculations are taken 
into account. However, the differences between NLO and NNLO nuclear 
PDFs are rather small for all other parton species. These findings 
are consistent with the perturbative convergence of the global 
$\chi^2$ discussed in Sec.~\ref{sec:minimizations} and listed in 
Tables~\ref{table:nuclear-DIS-data-D}, 
\ref{table:nuclear-DIS-data-C-Li}, 
\ref{table:Deuteron-structure-function}, 
\ref{table:neutrino-nucleus},
\ref{table:nuclear-DIS-data-JLAB} and
\ref{table:Drell-Yan}.
Concerning the fit quality of the 
total nuclear and neutrino DIS, and the Drell-Yan datasets, 
the most noticeable feature 
is a small improvement upon inclusion of higher-order QCD 
corrections. The inclusion of NNLO QCD corrections affects the 
nuclear PDFs uncertainty and improves the description of the data 
as well.

\subsection{ Comparison with other nuclear PDF sets }
\label{sec:Comparison-with-other-nuclear-PDFs-sets}

In this section, we present a comparison with the most 
recent nuclear PDF determinations available in the literature, 
namely {\tt nCTEQ15}~\cite{Kovarik:2015cma}, 
{\tt EPPS16}~\cite{Eskola:2016oht} and 
{\tt TUJU19}~\cite{Walt:2019slu}. Since the {\tt nCTEQ15} and 
{\tt EPPS16}  analyses were performed only at NLO accuracy, we 
limit the comparison to this perturbative order. All the 
comparisons presented in this section have been generated by 
using the standard {\tt LHAPDF6} library~\cite{Buckley:2014ana} 
and the published grids.

Each of these nuclear PDF analyses is based on a set of assumptions, 
for example, the form of the input parameterization at the initial
scale, the choice of the proton baseline PDFs, the included datasets 
and the kinematical cuts applied to the data, the perturbative 
order, and the scheme for the heavy quark contributions. 

In order to compare our results with other nuclear PDF sets, 
we begin with the detailed comparisons of nuclear modifications in lead.
In Fig.~\ref{fig:Modification-Model-TUJU-nCTEQ-EPPS-NLO}, we compare the 
{\tt KSASG20} nuclear modification factors 
in lead at NLO accuracy to those of 
{\tt nCTEQ15}~\cite{Kovarik:2015cma}, {\tt TUJU19}~\cite{Walt:2019slu} and {\tt EPPS16}~\cite{Eskola:2016oht} at 
$Q^2 = 100~{\rm GeV}^2$.
The comparison is presented per parton flavor $i$.
The bands for {\tt KSASG20} show 
the 68\% uncertainty estimation with $\Delta \chi^2 = 20$ obtained 
using the Hessian method.
One should keep in mind that the {\tt nCTEQ15}~\cite{Kovarik:2015cma} 
analysis is based on the tolerance criterion
$\Delta \chi^2 = 35$, while in the analysis by {\tt TUJU19}~\cite{Walt:2019slu} 
the condition $\Delta \chi^2 = 50$ is used 
and {\tt EPPS16}~\cite{Eskola:2016oht} have 
performed their error calculation for $\Delta \chi^2 = 52$. 
As one can see in
Fig.~\ref{fig:Modification-Model-TUJU-nCTEQ-EPPS-NLO}, in the case of valence 
up-quarks, we find a  behavior similar to {\tt nCTEQ15}
over the small and intermediate values of $x$,
but a stronger large-$x$ suppression.

In  the case of down-valence and gluon modifications, 
differences both in shape and uncertainty bands can 
be seen between  {\tt KSASG20}  and {\tt nCTEQ15}. 
The obtained valence modifications for {\tt KSASG20} are 
very similar both in shape and error bands.
For the case of strange-quark modification, the result is 
compatible with {\tt nCTEQ15} within
the estimated uncertainties at medium values of $x$.
A comparison of  {\tt KSASG20} with {\tt TUJU19} is also
presented in Fig.~\ref{fig:Modification-Model-TUJU-nCTEQ-EPPS-NLO}.
The up-valence and sea-quark modifications for  the
two fits can be considered
compatible since the error bands always overlap over the
whole range of $x$.
For the sea quark, the {\tt TUJU19} uncertainties appear
clearly larger than those of {\tt KSASG20}.
As can be seen, the central values for the gluon and down-valence are 
rather different. For both distributions, the {\tt TUJU19} uncertainties are 
clearly larger than
those of {\tt KSASG20}, except in the small-$x$ region for the gluon modification.
Our smaller uncertainties may be due to the fact that our 
parametrization is less flexible, especially in the case of 
the gluon density. 
In comparison with {\tt EPPS16}, we find several 
differences and similarities. As one can see from 
Fig.~\ref{fig:Modification-Model-TUJU-nCTEQ-EPPS-NLO},
the uncertainty for the sea-quark density for {\tt EPPS16} 
is much larger than the one of {\tt KSASG20}, and also larger 
than the results of other groups. For the gluon density we find 
compatible results at intermediate and small values of $x$, but 
{\tt KSASG20} and {\tt EPPS16} are different both in shape and 
central value at high-$x$. For the case of valence-quark nuclear 
modifications, one can again see that the central values are 
compatible only for $x<0.1$. The valence-quark uncertainty bands 
for {\tt KSASG20} are tighter than for {\tt EPPS16}.

In spite of some differences for the nuclear modification factors, when
calculating the total nuclear PDFs (see the next section),
we find good agreement with the other analyses available in the literature. 

For the full nuclear PDFs, we begin with a detailed comparison 
with the most recent nuclear 
PDF determination by {\tt TUJU19}. Regarding the experimental DIS 
data to determine the nuclear PDFs, both {\tt KSASG20} and 
{\tt TUJU19} are based on the same datasets with different 
kinematical cuts. However, in addition to the deuteron structure 
function $F_2^D$ from NMC~\cite{Arneodo:1996qe}, we also enrich 
our analysis with the $F_2^D$ data from 
BCDMS~\cite{Adams:1996gu,Benvenuti:1989fm} and 
HERMES~\cite{Airapetian:2011nu}, and data for the ratio 
$F_2^D / F_2^p$ from NMC~\cite{Arneodo:1996kd}.
Our {\tt KSASG20} analysis also incorporates the data 
from Drell-Yan cross-section ratios 
for several nuclear targets~\cite{Vasilev:1999fa,Alde:1990im} and 
the most recent DIS data from the
Jefferson Lab CLAS and Hall C experiments~\cite{Schmookler:2019nvf,Seely:2009gt}.
For the JLab data, target mass 
corrections and higher twist effects are taken into account 
in  {\tt KSASG20}.
The {\tt TUJU19} nuclear PDF sets are based on a {\tt CTEQ} proton 
baseline fitted within the same framework. {\tt TUJU19}  also 
assumed flavor symmetric sea quark densities, i.e. $\bar{u} = 
\bar{d} = s = \bar{s}$. The uncertainties for both analyses are 
obtained using the Hessian method, and {\tt TUJU19} calculated 
the uncertainty for $\Delta \chi^2 = 50$. 

The comparison with {\tt TUJU19} is presented in 
Fig.~\ref{fig:Compare-Model-TUJU-NNLO} at $Q^2 = 100$ GeV$^2$ 
for lead at NNLO accuracy. 
Concerning the shapes of these nuclear PDFs, a number of interesting 
differences between the two sets can be seen from the comparison 
presented in this figure. Small disagreements are found for the 
valence and sea-quark densities, however, the two sets still agree 
at the one-$\sigma$ level. A moderate difference is observed for 
the strange quark density below $x < 0.1$. A more pronounced 
difference in shape is observed for the gluon, charm and bottom 
quark PDFs, for which the {\tt KSASG20} distributions are more 
suppressed at medium to small values of $x$. The differences in 
shape among these three densities are more marked in the case of 
the gluon density and bottom quark PDFs. 
One should remember that neither of these 
analyses includes data which are directly sensitive to the gluon 
distribution. The origin of the differences between {\tt KSASG20} 
and {\tt TUJU19}, at medium to low $x$ for the gluon and sea-quark 
densities, and for medium $x$ in the case of valence quark PDFs, 
is likely to be due to  the input parameterization and 
a larger number of data points for the deuteron included in the 
{\tt KSASG20} analysis.
Another origin of these differences is due 
to the inclusion of Drell-Yan data and the most recent 
nuclear DIS data from JLab in our analysis.

In Figs.~\ref{fig:Compare-Model-TUJU-nCTEQ-EPPS-NLO-1} and 
\ref{fig:Compare-Model-TUJU-nCTEQ-EPPS-NLO-2}, our nuclear 
PDFs at the scale $Q^2 = 100$ GeV$^2$ are presented for lead at 
NLO accuracy. The most recent NLO nuclear PDF determinations 
available in the literature, namely from  
{\tt nCTEQ15}~\cite{Kovarik:2015cma}, 
{\tt EPPS16}~\cite{Eskola:2016oht} and 
{\tt TUJU19}~\cite{Walt:2019slu} are also shown for comparison. 
We should mention here that the {\tt nCTEQ15} analysis is based 
on the tolerance criterion $\Delta \chi^2 = 35$, while {\tt EPPS16} 
presented their results for $\Delta \chi^2 = 52$. 
The uncertainty bands for the {\tt KSASG20} nuclear PDFs 
are obtained using the Hessian method with  $\Delta \chi^2 = 20$,
and are related to the parameters in nuclear modification 
factors.

In the {\tt EPPS16} analysis, the bound nucleon PDFs are defined 
relative to the free nucleon baseline {\tt CT14} 
PDFs~\cite{Dulat:2015mca}, as for the case of our previous 
study {\tt KA15}~\cite{Khanpour:2016pph} where we considered the same framework
and used {\tt JR} PDFs~\cite{JimenezDelgado:2008hf}. 
The EPPS16 analysis 
was the first study which used data from the LHC for $Z$ and $W^\pm$ 
boson~\cite{Khachatryan:2015hha,Khachatryan:2015pzs, 
Aad:2015gta} and dijet production~\cite{Chatrchyan:2014hqa} in 
proton-lead collisions. These collider data provide further 
constraints for the gluon nuclear modifications and for the flavor 
separation. In the {\tt nCTEQ15} analysis, the nuclear PDFs are 
parameterized by a polynomial functional form, in which all the 
$A$ dependence is encoded in the coefficients of the 
parameterization. {\tt nCTEQ15} assumed $s = \bar{s}$ and 
the strange quark density is related to $\bar{u} + \bar{d}$ 
by an additional $A$-dependent factor, $s = \bar{s} = 
(\kappa(A)/2) (\bar{u} + \bar{d})$. The {\tt TUJU19} analysis 
considered flavor symmetry for the sea quark densities, $\bar{u} 
= \bar{d} = s = \bar{s}$, while in {\tt EPPS16} and {\tt KSASG20} 
only  $s = \bar{s}$ was used as a constraint. 

After presenting the main properties of these recent nuclear PDFs, 
we compare their results at NLO accuracy. As can be seen from 
Fig.~\ref{fig:Compare-Model-TUJU-nCTEQ-EPPS-NLO-1}, for the $xu_v$ 
and $xd_v$ densities, the {\tt KSASG20} results are in agreement 
in size with {\tt TUJU19}, {\tt nCTEQ15} and {\tt EPPS16}, and 
well within their uncertainties, despite differences in the dataset 
and parametrization, etc.
For the valence quark densities we find that both {\tt KSASG20} 
$xu_v$ and $xd_v$ slightly tend to stay below all other results 
at medium values of $x \sim 0.1$. Moderate differences 
for the gluon density and significant differences for the sea-quark 
densities are observed. For both cases, {\tt EPPS16} exhibits 
relatively wider error bands compared with other analyses. The 
gluon distributions from {\tt nCTEQ15} and {\tt EPPS16} fall 
below the one of {\tt KSASG20}. 

In Fig.~\ref{fig:Compare-Model-TUJU-nCTEQ-EPPS-NLO-2} we show a 
comparison for the sea-quark distributions (upper row) and the 
charm and the bottom distributions (lower row). The latter two 
are perturbatively generated.
Moderate differences between {\tt KSASG20}, {\tt nCTEQ15}  
{\tt TUJU19}, and {\tt EPPS16} can be seen. 
We find that the agreement between the 
results of {\tt KSASG20} and {\tt EPPS16} for the central 
values of $x\bar{u}$, $x\bar{d}$ and $x\bar{c}$ for the whole 
range of $x$ is slightly better than with the other two groups.
The resulting 
uncertainties for {\tt KSASG20} are somewhat smaller. 
We should stress again that all results presented here 
are based on the choice $\Delta \chi^2 = 20$ for the tolerance.  
Choosing the larger tolerance value $\Delta \chi^2 = 50$, as 
preferred by other groups, the error bands of our nuclear PDFs 
would increase by a factor of $\sim 1.5$.

\subsection{ Fit quality and comparison of data and theory } 
\label{sec:Fit-quality}

In Tables~\ref{table:nuclear-DIS-data-D}, 
\ref{table:nuclear-DIS-data-C-Li}, 
\ref{table:Deuteron-structure-function}, 
\ref{table:neutrino-nucleus},
\ref{table:nuclear-DIS-data-JLAB}, 
\ref{table:Drell-Yan} and
\ref{tab-nuclear-pdf}  
presented above in section~\ref{sec:Nucleardata} we have shown 
the values of $\chi^2$ per data point for each 
individual dataset, both for the NLO and the NNLO fits. 

In total, we find for the {\tt KSASG20} fit at NLO, $\chi^2 = 
4582$ with 4353 data points. 
With 18 free parameters this 
leads to a $\chi^2/{\rm d.o.f} = 1.05$  which indicates a 
relatively good fit. At NNLO, the {\tt KSASG20} fit leads to 
$\chi^2/{\rm d.o.f} = 1.04$. This is only a moderate improvement 
upon inclusion of the higher-order QCD corrections.

While most datasets for nuclear and neutrino DIS experiments
and Drell-Yan data
satisfy the goodness of fit criterion, there are some experiments 
which stand out as having a poor fit. We also notice that for some 
datasets, $\chi^2$ is poor even for the NNLO fit. In addition, 
for some individual datasets $\chi^2$ increases as higher-order QCD 
corrections are included. A similar observation was made in 
previous analyses by {\tt TUJU19}~\cite{Walt:2019slu} and 
{\tt nNNPDF1.0}~\cite{AbdulKhalek:2019mzd}.

As one can see from Table~\ref{table:neutrino-nucleus}, with 
the exception of $\nu$Pb data from CHORUS, for all other 
(anti)neutrino-nucleus DIS data the inclusion of higher-order 
QCD corrections leads to a better fit quality. 

In order to illustrate our discussion, we present selected 
comparisons of the datasets used in this study to the corresponding 
NNLO theoretical predictions obtained using the {\tt KSASG20} NNLO 
nuclear PDFs. 
In the following plots, we combine experimental data for each 
nucleus for a wider range of $Q^2$-values (roughly between 1.8 
and 67~GeV$^2$) where scaling violations are observed 
to cancel in the considered ratio. Our NNLO results 
have been calculated at $Q^2 = 5$~GeV$^2$. 
In Fig.~\ref{fig:Ratio-F2A-over-F2C} such a comparison 
is displayed for the nuclear DIS data for the ratio $F_2^A(x, Q^2) / 
F_2^{C} (x, Q^2)$ as a function of $x$.
In Fig.~\ref{fig:Ratio-F2A-over-F2D} we also compare our NNLO 
results for $F_2^A(x, Q^2) / F_2^{D} (x, Q^2)$ and 
$F_2^A(x, Q^2) / F_2^{Li} (x, Q^2)$ as a function of $x$ with 
some selected nuclear DIS data.  The data 
shown in these plots are measured by the NMC and E139 
Collaborations. The bands show the 68\% uncertainty estimates 
with $\Delta \chi^2 = 20$.
The comparisons presented in these plots demonstrate that the 
agreement between our NNLO theoretical predictions and the 
nuclear DIS experimental measurements varies between different 
data for different nuclei. Apart from a few data points in the 
small-$x$ region, the agreement with most of the data published 
by the NMC and E139 Collaborations is excellent.

A detailed comparison with the 
CHORUS data on neutrino(antineutrino) lead 
collisions~\cite{Onengut:2005kv} and the CDHSW data on 
neutrino(antineutrino) iron collisions~\cite{Berge:1989hr} are 
shown in Figs.~\ref{fig:Neutrino-Antineutrino-Lead} and 
\ref{fig:Neutrino-Antineutrino-Iron}, respectively.
The results are shown as a function of $Q^2$ for some selected 
bins of $x$ and $y$. The incident beam of neutrino(antineutrino) 
energies are not high enough to reach small values of $x$. Here  
we consider the range from $x = 0.125$ to $x=0.65$, corresponding 
to the range between $y=0.3$ and $y=0.5$. Very good agreement is 
achieved for the neutrino(antineutrino)-nucleus data presented 
in these plots for the whole region of $x$ and $Q^2$.

The data from JLab experiment~\cite{Schmookler:2019nvf} included 
in this study provide important additional constraints on the 
shape and the uncertainty bands of nuclear PDFs in the high-$x$
and low-Q$^2$ regime~\cite{Paukkunen:2020rnb}. In order to show 
the agreement between our results using the 
extracted nuclear PDFs, in Fig.~\ref{fig:F2-A-F2-D-CLAS-Final},
we present the data/theory comparison for carbon (C), aluminum 
(Al), iron (Fe), and lead (Pb) nuclei. The bands in this plot 
show the 68\% uncertainty estimates with $\Delta \chi^2 = 20$.
We have again combined experimental data for different 
$Q^2$-values (between 1.8 and 46~GeV$^2$) while our theory 
results are calculated at $Q^2$ = 3 GeV$^2$.
As one can see, our NLO and NNLO theory results nicely 
describe the recent JLab CLAS~\cite{Schmookler:2019nvf} data.

We now turn to the Drell-Yan cross section ratios 
$\sigma_{{\text{DY}}}^A/\sigma_{{\text{DY}}}^{A^{\prime}}$.
In Fig.~\ref{fig:DY}, we display the cross section ratio measured 
by the Fermilab experiment E866 for some selected nuclear 
targets~\cite{Vasilev:1999fa}. 
Our NLO and NNLO results along with their uncertainties are 
shown as well. The Drell-Yan cross sections 
are presented in four bins of the invariant mass
and are shown as a function of the 
proton momentum fraction $x_{1}$.
As one can see, except for some isolated points 
with relatively large errors, 
the {\tt KSASG20} NLO theory results describe the data well.
In addition, the results show that the uncertainty bands 
at NNLO are slightly smaller than at NLO.

%
\section{Summary and conclusions} 
\label{sec:Discussion}
%

In summary, in this work, we have introduced a new set of nuclear 
PDFs at NLO and NNLO accuracy in pQCD. 
{\tt KSASG20} nuclear 
PDFs are obtained from most up-to-date experimental data, including 
neutral-current nuclear DIS with several nuclear targets, charged-current 
neutrino DIS, and Drell-Yan cross-section measurements. The combination 
of these data are sensitive to the flavor decomposition of nuclei.
Our analysis 
also incorporates the most recent DIS data from the Jefferson Lab CLAS 
and Hall C experiments. For these specific data sets, we take into 
account target mass corrections and higher twist effects which are 
mainly important in the region of large $x$ and intermediate-to-low Q$^2$. 
Heavy quark mass effects are included in the {\tt FONLL} 
general-mass variable-flavor-number scheme for charm and bottom quarks. 

Our determination of nuclear PDFs includes error estimates 
obtained within the Hessian method with the tolerance criterion 
of $\Delta \chi^2=20$. The effects arising from the inclusion 
of higher-order QCD corrections are investigated. 
We found only small differences between the NLO and NNLO QCD 
fits, both for the shape and size of the uncertainty bands. 
The inclusion of higher-order QCD corrections slightly improves 
the description of the nuclear data analyzed in this study, but 
there is no strong indication that NNLO corrections are required 
by the data. The largest differences appear for the gluon nuclear 
PDF, where our NNLO fit leads to a slight decrease in the 
uncertainties. 

In contrast to other recent analyses, such as 
{\tt nNNPDF2.0}~\cite{AbdulKhalek:2019mzd} and 
{\tt nNNPDF2.1}~\cite{AbdulKhalek:2020yuc} which use Monte 
Carlo techniques based on the NNPDF framework, and 
{\tt TUJU19}~\cite{Walt:2019slu} which is based on the CTEQ 
framework, we parameterize the nuclear PDFs considering 
nuclear correction factors. To this end, a 
modern set of parton distribution functions for free protons, 
namely {\tt CT18}, is considered as reference. Our results 
are consistent, within uncertainties, with the previous 
determinations of nuclear PDFs available in the literature, 
in particular for {\tt nCTEQ15}~\cite{Kovarik:2015cma}, 
{\tt EPPS16}~\cite{Eskola:2016oht} and {\tt TUJU19}, which are 
based on a different selection of data sets
and assumptions. However, we found 
a number of differences which only occur in regions without any 
constraints from data. These differences can be attributed to 
different assumptions such as the input parametrization of the 
nuclear modification factors.

The nuclear PDFs analysis presented in this article represents 
the first step of a broader study. A number of improvements
are foreseen for the near future.
While the data used in the present study allowed us to determine 
the nuclear quark and anti-quark densities, the nuclear gluon PDF 
is only loosely constrained. This is actually the most important 
limitation of the {\tt KSASG20} nuclear PDFs analysis. 
To resolve this limitation, we plan to include 
additional datasets, especially from present and future 
measurements of proton-lead and lead-lead collisions at the 
CERN-LHC, which are expected to provide direct information on the 
nuclear gluon modifications and more stringent flavor-dependence 
constraints. 
These include, for example, the data from PHENIX and STAR for 
inclusive pion production in deuteron-gold (d-Au) 
collisions~\cite{Adler:2006wg,Abelev:2009hx}, neutral pion 
production data from PHENIX~\cite{Adler:2006wg}, and the charged 
and neutral pion data from the STAR 
experiment~\cite{Adams:2006nd,Abelev:2009hx}. 
Also data from proton-lead (p-Pb) collisions from the ATLAS 
and CMS Collaborations at the LHC, which have been included 
in the analyses of nuclear PDFs performed in 
Refs.~\cite{AbdulKhalek:2020yuc,Eskola:2016oht}, are expected to 
improve the constraints of the nuclear gluon density at large 
momentum fractions. With more precise data, it might be necessary 
in the future to consider more flexible parametrizations. 

We can also expect that the electron-ion collider 
(EIC)~\cite{Accardi:2012qut}, the Large Hadron-Electron Collider 
(LHeC)~\cite{Agostini:2020fmq} or a Future Circular Collider 
(FCC)~\cite{Abada:2019lih,Benedikt:2018csr} could provide precise 
data for nuclear PDF analyses. It would 
be very interesting to repeat the analysis described here using 
additional observables from hadron colliders such as the LHC.

The NLO and NNLO nuclear PDF sets presented in this work are 
available in the {\tt LHAPDF} format~\cite{Buckley:2014ana} for 
all relevant nuclei from $A = 2$ to $A = 208$ and can be obtained 
from the authors upon request.

\section*{Acknowledgments}

We are thankful to Petja Paakkinen, Hamed Abdolmaleki, 
Marina Walt, Ilkka Helenius and Muhammad Goharipour for many 
helpful discussions and comments.
We also gratefully thank 
Valerio Bertone for a discussion 
regarding subtle differences between various heavy-quark mass 
schemes and their numerical implementation.  
The authors thank the School of Particles and Accelerators, 
Institute for Research in Fundamental Sciences (IPM) for financial 
support of this project. Hamzeh Khanpour also is grateful to the 
University of Science and Technology of Mazandaran for financial 
support of this project.

\clearpage

%

\end{document}